\documentclass[twocolumn, trackchanges]{aastex631}
\usepackage{amsmath}
\usepackage{CJKutf8}
\newcommand{\molhy}{H$_2$}
\newcommand{\um}{$\mu$m}
\newcommand{\feii}{[Fe \textsc{ii}]}
\newcommand{\paa}{Pa$\alpha$}
\newcommand{\pab}{Pa$\beta$}

\newcommand{\pad}{Pa$\delta$}
\newcommand{\brb}{Br$\beta$}
\newcommand{\brg}{Br$\gamma$}
\newcommand{\brd}{Br$\delta$}

\newcommand{\sivii}{[Si \textsc{vii}]}
\newcommand{\mgviii}{[Mg \textsc{viii}]}

\accepted{\today}

\shorttitle{GOALS-JWST NGC\,7469 NIRSpec}
\shortauthors{Bianchin et al.}

\begin{document}
\begin{CJK*}{UTF8}{bsmi}

\title{GOALS-JWST: Gas Dynamics and Excitation in NGC\,7469 revealed by NIRSpec}

\author[0000-0002-6570-9446]{Marina Bianchin}
\altaffiliation{IAU-Gruber Foundation Fellow} 
\affiliation{4129 Frederick Reines Hall, Department of Physics and Astronomy, University of California, Irvine, CA 92697, USA}\email{mbianch1@uci.edu}

\author[0000-0002-1912-0024]{Vivian U}
\affiliation{4129 Frederick Reines Hall, Department of Physics and Astronomy, University of California, Irvine, CA 92697, USA} 

\author[0000-0002-3139-3041]{Yiqing Song}
\affiliation{European Southern Observatory, Alonso de Córdova, 3107, Vitacura, Santiago, 763-0355, Chile}
\affiliation{Joint ALMA Observatory, Alonso de Córdova, 3107, Vitacura, Santiago, 763-0355, Chile}

\author[0000-0001-8490-6632]{Thomas S.-Y. Lai (賴劭愉)}
\affil{IPAC, California Institute of Technology, 1200 E. California Blvd., Pasadena, CA 91125}

\author[0000-0002-0164-8795]{Raymond P. Remigio}
\affiliation{4129 Frederick Reines Hall, Department of Physics and Astronomy, University of California, Irvine, CA 92697, USA}
\author[0000-0003-0057-8892]{Loreto Barcos-Mu\~noz}
\affiliation{National Radio Astronomy Observatory, 520 Edgemont Road, Charlottesville, VA, 22903, USA}
\affiliation{Department of Astronomy, University of Virginia, 530 McCormick Road, Charlottesville, VA, 22903, USA}

\author[0000-0003-0699-6083]{Tanio D\'iaz-Santos}
\affiliation{Institute of Astrophysics, Foundation for Research and Technology-Hellas (FORTH), Heraklion, 70013, Greece}
\affiliation{School of Sciences, European University Cyprus, Diogenes street, Engomi, 1516 Nicosia, Cyprus}

\author[0000-0003-3498-2973]{Lee Armus}
\affil{IPAC, California Institute of Technology, 1200 E. California Blvd., Pasadena, CA 91125}
\author[0000-0003-4268-0393]{Hanae Inami}
\affiliation{Hiroshima Astrophysical Science Center, Hiroshima University, 1-3-1 Kagamiyama, Higashi-Hiroshima, Hiroshima 739-8526, Japan}

\author[0000-0003-3917-6460]{Kirsten L. Larson}
\affiliation{AURA for the European Space Agency (ESA), Space Telescope Science Institute, 3700 San Martin Drive, Baltimore, MD 21218, USA}

\author[0000-0003-2638-1334]{Aaron S.~Evans}
\affiliation{National Radio Astronomy Observatory, 520 Edgemont Road, Charlottesville, VA, 22903, USA}
\affiliation{Department of Astronomy, University of Virginia, 530 McCormick Road, Charlottesville, VA, 22903, USA}

\author[0000-0002-5666-7782]{Torsten B\"oker}
\affiliation{European Space Agency, c/o STScI, 3700 San Martin Drive, Baltimore, MD 21218, USA}

\author[0000-0002-6650-3757]{Justin A. Kader}
\affiliation{4129 Frederick Reines Hall, Department of Physics and Astronomy, University of California, Irvine, CA 92697, USA}

\author[0000-0002-1000-6081]{Sean T.~Linden}
\affiliation{Department of Astronomy, University of Massachusetts at Amherst, Amherst, MA 01003, USA}
\author[0000-0002-2688-1956]{Vassilis Charmandaris}
\affiliation{Institute of Astrophysics, Foundation for Research and Technology-Hellas (FORTH), Heraklion, 70013, Greece}
\affiliation{School of Sciences, European University Cyprus, Diogenes street, Engomi, 1516 Nicosia, Cyprus}
\affiliation{Department of Physics, University of Crete, Heraklion, 71003, Greece}

\author[0000-0001-6919-1237]{Matthew A. Malkan}
\affiliation{Department of Physics \& Astronomy, 430 Portola Plaza, University of California, Los Angeles, CA 90095, USA}
\author[0000-0002-5807-5078]{Jeff Rich}
\affiliation{The Observatories of the Carnegie Institution for Science, 813 Santa Barbara Street, Pasadena, CA 91101}

\author[0000-0002-4375-254X]{Thomas Bohn}
\affil{Hiroshima Astrophysical Science Center, Hiroshima University, 1-3-1 Kagamiyama, Higashi-Hiroshima, Hiroshima 739-8526, Japan}

\author[0000-0001-7421-2944]{Anne M.~Medling}
\affiliation{Department of Physics \& Astronomy and Ritter Astrophysical Research Center, University of Toledo, Toledo, OH 43606,USA}
\affiliation{ARC Centre of Excellence for All Sky Astrophysics in 3 Dimensions (ASTRO 3D), Australia}

\author[0000-0002-2596-8531]{Sabrina Stierwalt}
\affiliation{Physics Department, 1600 Campus Road, Occidental College, Los Angeles, CA 90041, USA}
\author[0000-0002-8204-8619]{Joseph M. Mazzarella}
\affiliation{IPAC, California Institute of Technology, 1200 E. California Blvd., Pasadena, CA 91125}

\author[0000-0002-9402-186X]{David R.~Law}
\affiliation{Space Telescope Science Institute, 3700 San Martin Drive, Baltimore, MD 21218, USA}

\author[0000-0003-3474-1125]{George C. Privon}
\affiliation{National Radio Astronomy Observatory, 520 Edgemont Road, Charlottesville, VA, 22903, USA}
\affiliation{Department of Astronomy, University of Virginia, 530 McCormick Road, Charlottesville, VA, 22903, USA}
\affiliation{Department of Astronomy, University of Florida, Gainesville, FL 32611, USA}

\author[0000-0002-5828-7660]{Susanne Aalto}
\affiliation{Department of Space, Earth and Environment, Chalmers University of Technology, 412 96 Gothenburg, Sweden}
\author{Philip Appleton}
\affiliation{IPAC, California Institute of Technology, 1200 E. California Blvd., Pasadena, CA 91125}

\author{Michael J.~I. Brown}
\author{Victorine A. Buiten}
\affiliation{Leiden Observatory, Leiden University, PO Box 9513, 2300 RA Leiden, The Netherlands}

\author[0000-0002-1392-0768]{Luke Finnerty}
\affiliation{Department of Physics \& Astronomy, 430 Portola Plaza, University of California, Los Angeles, CA 90095, USA}

\author[0000-0003-4073-3236]{Christopher C. Hayward}
\affiliation{Center for Computational Astrophysics, Flatiron Institute, 162 Fifth Ave., New York, NY 10010, USA}

\author[0000-0001-6028-8059]{Justin Howell}
\affil{IPAC, California Institute of Technology, 1200 E. California Blvd., Pasadena, CA 91125}

\author[0000-0002-4923-3281]{Kazushi Iwasawa}
\affiliation{ICREA, Pg. Llu\'is Companys 23, 08010 Barcelona, Spain}
\affiliation{Institut de Ci\`encies del Cosmos (ICCUB), Universitat de Barcelona (IEEC-UB), Mart\'i i Franqu\`es, 1, 08028 Barcelona, Spain}

\author[0000-0003-2743-8240]{Francisca Kemper}
\affiliation{Institut de Ci\`encies de l'Espai (ICE, CSIC), Can Magrans, s/n, E-08193 Cerdanyola del Vall\`es, Barcelona, Spain}
\affiliation{ICREA, Pg. Llu\'is Companys 23, 08010 Barcelona, Spain}
\affiliation{Institut d'Estudis Espacials de Catalunya (IEEC), E-08034 Barcelona, Spain}


\author{Jason Marshall}
\affiliation{Glendale Community College, 1500 N. Verdugo Rd., Glendale, CA 91208}

\author{Jed McKinney} 
\affiliation{Department of Astronomy, University of Massachusetts at Amherst, Amherst, MA 01003, USA}

\author[0000-0002-2713-0628]{Francisco M\"uller-S\'anchez}
\affiliation{Department of Physics and Materials Science, The University of Memphis, 3720 Alumni Avenue, Memphis, TN 38152, USA}

\author[0000-0001-7089-7325]{Eric J.\,Murphy}
\affiliation{National Radio Astronomy Observatory, 520 Edgemont Road, Charlottesville, VA, 22903, USA}

\author[0000-0001-5434-5942]{Paul P. van der Werf}
\affiliation{Leiden Observatory, Leiden University, PO Box 9513, 2300 RA Leiden, The Netherlands}

\author[0000-0002-1233-9998]{David B. Sanders}
\affiliation{Institute for Astronomy, University of Hawaii, 2680 Woodlawn Drive, Honolulu, HI 96822, USA}
\author[0000-0001-7291-0087]{Jason Surace}
\affiliation{IPAC, California Institute of Technology, 1200 E. California Blvd., Pasadena, CA 91125}

\begin{abstract}
We present new JWST-NIRSpec IFS data for the luminous infrared galaxy NGC\,7469: a nearby (70.6\,Mpc) active galaxy with a Sy 1.5 nucleus that drives a highly ionized gas outflow and a prominent nuclear star-forming ring.
Using the superb sensitivity and high spatial resolution of the {\em JWST} instrument NIRSpec-IFS, we investigate the role of the Seyfert nucleus in the excitation and dynamics of the circumnuclear gas. Our analysis focuses on the \feii, \molhy, and hydrogen recombination lines that trace the radiation/shocked-excited molecular and ionized ISM around the AGN. We investigate the gas excitation through \molhy/\brg~and \feii/\pab~emission line ratios and find that photoionization by the AGN {dominates 
within the central 300\,pc of the galaxy except in a small region that shows signatures of shock-heated gas}; these shock-heated regions are likely associated with a compact radio jet.  In addition, the velocity field and velocity dispersion maps reveal complex gas kinematics. Rotation is the dominant feature, but we also identify non-circular motions consistent with gas inflows as traced by the velocity residuals and the spiral pattern in the \paa~velocity dispersion map. 
The inflow is 
two orders of magnitude higher than the AGN accretion rate. The compact nuclear radio jet has enough power to drive the highly ionized outflow. This scenario suggests that the inflow and outflow are in a self-regulating feeding-feedback process, with a contribution from the radio jet helping to drive the outflow. 



\end{abstract}

\section{Introduction} 
Luminous infrared galaxies (LIRGs) are systems with high infrared luminosities ($L_{\rm IR}\geq10^{11}L_{\odot}$) that, in the nearby universe, are mostly major mergers \citep{sanders96}. These galaxies host extreme environments where both starburst and nuclear activity play a significant role in shaping their evolution \citep[e.g.][]{hekatelyne20}. Spatially resolved studies show that the gas excitation in LIRGs and their ultraluminous counterparts (ULIRGs, $L_{\rm IR}\geq10^{12}L_{\odot}$) is governed by a mixture of processes typical of active galactic nuclei \citep[AGN, sometimes hidden behind large amounts of dust with $A_V$ up to $\sim40$;][]{u19,perez-torres21}, star-forming regions, and shocked gas \citep[e.g.][]{rich15, hekatelyne20}.

The gas dynamics in (U)LIRGs are often heavily impacted by the interaction with other galaxies, the presence of AGNs and/or Starbursts. Evidence for this comes from the detection of outflows on a variety of physical scales and gas phases: cold \citep{feruglio10,xu14, veilleux17, falstad18, barcos-munoz18,lutz20, pereira-santaella20} and hot molecular gas \citep{u13,u19,medling15,riffel20, motter21}, as well as moderately- \citep{rich12,rich14, arribas14, rich15,robleto-orus21,xu22} and highly-ionized gas \citep{rodriguez-ardila06,muller-sanchez11,u22c,armus23} and neutral gas \citep{rupke11,morganti16,su23}.  Streaming motions, such as gas inflows, toward the nuclear regions of ULIRGs have also been detected in some sources \citep{hekatelyne18, medling19, aalto19, gonzalez-alfonso21}.  These inflows may act in providing the fuel necessary to trigger and feed the AGN.

\begin{figure*}
    \centering
    \includegraphics[width=\textwidth,trim={14cm 0 0.5cm 0},clip]{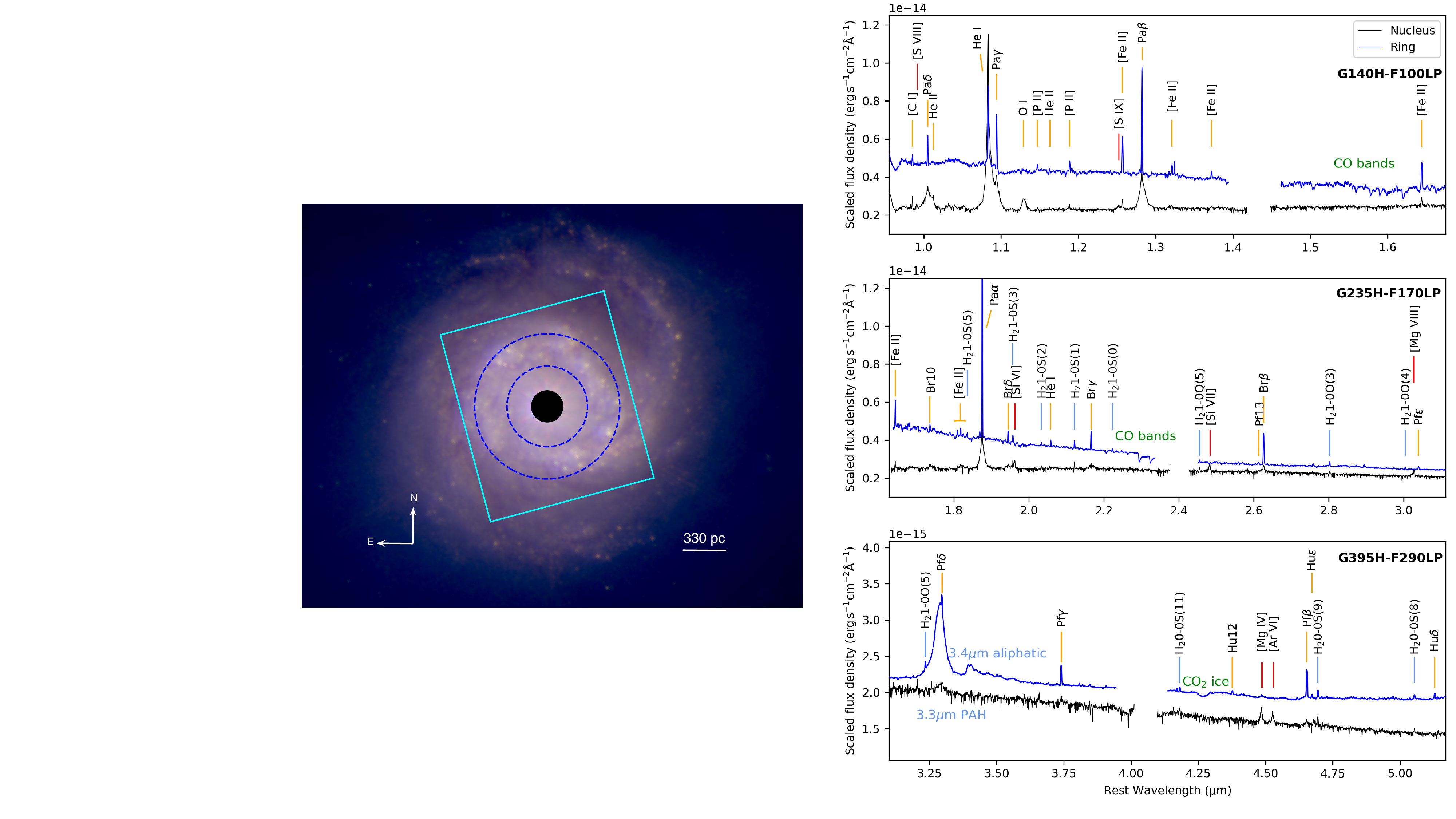}
    \caption{\emph{Left: HST} (F110W) and \emph{JWST} (F150W and F200W) color composite image of NGC~7469 \citep{bohn23}. North is up and East to the left as indicated by the compass in the bottom left. 
    Bright stellar clusters are detected throughout the starburst ring
    The cyan rectangle indicates the combined FoV of our dithered NIRSpec observations ($4.2\arcsec\times4.8\arcsec$). 1$\arcsec$ corresponds to 330\,pc at the distance of the galaxy. The right panels show extracted NIRSpec spectra from the nucleus with a radius of 0.4\,arcsec (black) and the star formating ring with inner and outer radius of 1~arcsec and 1.8~arcsec (blue). These extractions correspond to the black-filled circle and the dashed blue annulus in the galaxy image. The red, orange, and light blue lines indicate the location of the high ionization (coronal), the ionized, and molecular emission lines, respectively. {The spectra were arbitrarily scaled for visualization purposes to differentiate between the nuclear and the star-forming ring spectra.} The first shows the broad lines, typical of Type 1 nuclei, in the permitted emission lines as the Hydrogen and Helium recombination lines, high ionization coronal lines indicated by the red markers, and other emission lines as \feii~and \molhy. The latter is dominated by narrow lines, a prominent PAH feature at $3.3\mu$m and the aliphatic at 3.4$\mu$m, CO stellar absorption features, and a CO$_2$ absorption.}
    \label{fig:overview}
\end{figure*}

{NGC\,7469 hosts a Type I active nucleus \citep{osterbrock77, peterson14} sub-classified as Seyfert 1.5 }
\citep{landt08} located at 70.6 Mpc ($z=0.01627$) with a (R')SAB(rs)a morphology \citep{RC3}. The central kiloparsec region of NGC\,7469 shows a bright AGN surrounded by a star-forming ring \citep[e.g.][]{davies04,diaz-santos07,song21}. The ring has inner and outer radii of 330\,pc and 616\,pc (see Fig.~\ref{fig:overview}) with a bimodal stellar population as seen from HST imaging \citep{diaz-santos07} with a star formation rate of $\sim22$\,M$_{\odot}$yr$^{-1}$ estimated from the radio continuum \citep{song21} consistent with $10-30$\,M$_{\odot}$yr$^{-1}$ estimated from the recombination, [Ne{\sc ii}], and [Ne{\sc iii}] \citep{lai22}. \emph{JWST} NIRCam \citep{rieke23} and MIRI \citep{bouchet15} imaging reveal previously undetected embedded stellar clusters with colors that suggest stellar ages of $<5$\,Myr \citep{bohn23}. The H$_2$/PAH ratio, which measures the excess \molhy~relative to PAHs, is within the range of values from normal photo-dissociation models implying that the ring is not currently being affected by AGN feedback \citep{lai22}. 

NGC\,7469 hosts a nuclear outflow traced by highly ionized gas as evidenced in the \emph{JWST} MIRI MRS observations 
\citep{u22c,armus23} and in the Near-IR via the {[Si\,{\sc vi}] $1.96\mu$m emission line} \citep{muller-sanchez11}, {and by the moderately ionized gas also traced by MIRI MRS \citep{zhang23}. This outflow has an extent of 400-600\,pc} and affects mostly the region located between the nucleus and the ring, referred to as the inner interstellar medium (inner-ISM). The outflow has a stratified and decelerating structure as evidenced by highly blueshifted wings on the coronal emission lines 
\citep{armus23}. 
The warm molecular gas shows enhanced velocity dispersion to the northwest of the central AGN, suggestive of the presence of shocked gas \citep{u22c}. Considering the richness of physical phenomena in its inner kiloparsec region, NGC\,7469 is an ideal target to investigate the AGN--starburst--ISM interaction.

In this paper, we report the morphology, kinematics, and excitation of the ionized atomic and hot ($\sim 1000$~K) molecular gas in the central region of NGC\,7469 as observed by \emph{JWST} \citep[and references therein]{rigby23} with the  NIRSpec instrument in its integral field spectroscopy (IFS) mode. 
We describe the observations and data reduction in Section \ref{sec:obs} and our results in Section \ref{sec:results}. We discuss the kinematics and gas excitation in Section \ref{sec:discussion} and present our conclusions in Section \ref{sec:conclusions}.

\section{Data}
\label{sec:obs}
\subsection{\emph{JWST}-NIRSpec Observations}

NGC\,7469 was observed with the NIRSpec IFS \citep{boker22,jakobsen22} on July 19th 2022 UT as part of the \emph{JWST} Director Discretionary Time Early Release Science (DD-ERS) program (PID: 1328, PIs Lee Armus \& Aaron Evans). We employed NIRSpec in the high resolution mode ($R\approx2700$, { corresponding to a velocity resolution of $\approx110$\,km\,s$^{-1}$) in three grating/filter combinations --- G140H/F100LP, G235H/F170LP, G395H/F290LP --- covering the wavelength range of 0.97 to 5.27$\mu$m with a nominal field of view (FoV) of $3\arcsec \times3$\arcsec~(see Figure \ref{fig:overview}).} In order to fully cover the star-forming ring ($\sim 4\arcsec$ across), we used the large cycling four-point dither pattern, which provides a FoV of $4.2\arcsec\times4.8\arcsec$ corresponding to $1.4\times1.6$\,kpc$^2$ at the distance of NGC\,7469 (Fig.~\ref{fig:overview}). 
For extended targets, MSA leakage correction (leakcal) exposures are necessary to account for the leakage from the permanently open micro-shutters. While observing NGC\,7469, we used the same dither pattern for the leakcals as we did for the science observations.  The science exposure times in each grating/filter combination is 817~s and the total observing time, including the leakcals and overheads, is 10.5~ks.

\subsection{Data reduction}
We reduce the level-1 data downloaded from MAST using the JWST pipeline version 1.8.3 \citep{bushouse2022} in batch mode.  The reference files follow the context version \texttt{jwst\_12027.pmap}. The first reduction step is \texttt{Detector1}, which generates rate files with detector level correction applied to them. 
Both science and leakage calibration files are processed at the \texttt{Detector1} step. The second reduction step for spectroscopic data is \texttt{Spec2}, which applies the distortion, wavelength, and flux calibrations, and other 2D corrections to the science data, including the leakage subtraction.

The outlier rejection algorithm is not efficient in the current version of the pipeline. Most of the outliers have fluxes higher than the typical flux of the brighter emission lines or show negative flux values. We optimized the outlier rejection by flagging spectral pixels that showed clear contamination in the calibrated science frames {\citep{hutchison23}}. These corrected frames are then fed to the last reduction step \texttt{Spec3} that builds the data cubes by combining individual exposures taken at each dither position. We obtain three data cubes corresponding to each grating and filter combination: G140H/100LP, G235H/170LP and G395H/290LP. 

In order to correct the pipeline-processed cubes for astrometry we adopt the peak of the continuum in each cube as the position of the nucleus. We then modify the NIRSpec-IFS data cube headers to align with the continuum peak of CH1-short MIRI-MRS cube.  
Our final datacubes have a {spatial resolution of $0.14\arcsec$ at $1.12\mu$m as obtained from the FWHM of the O\,{\sc i}\,$1.12 \mu$m emission line image, allowing us to resolve structures down to $\sim 50$\,pc at this wavelength. The spatial resolution is poorer at longer wavelengths, as demonstrated by \citet{lai23} that measure a spatial resolution of $0.19\arcsec$ for the dust features at 3.3$\mu$m.
}


\begin{figure*}
    \centering
    \includegraphics[width=\textwidth,trim=35 0 20 35,clip]{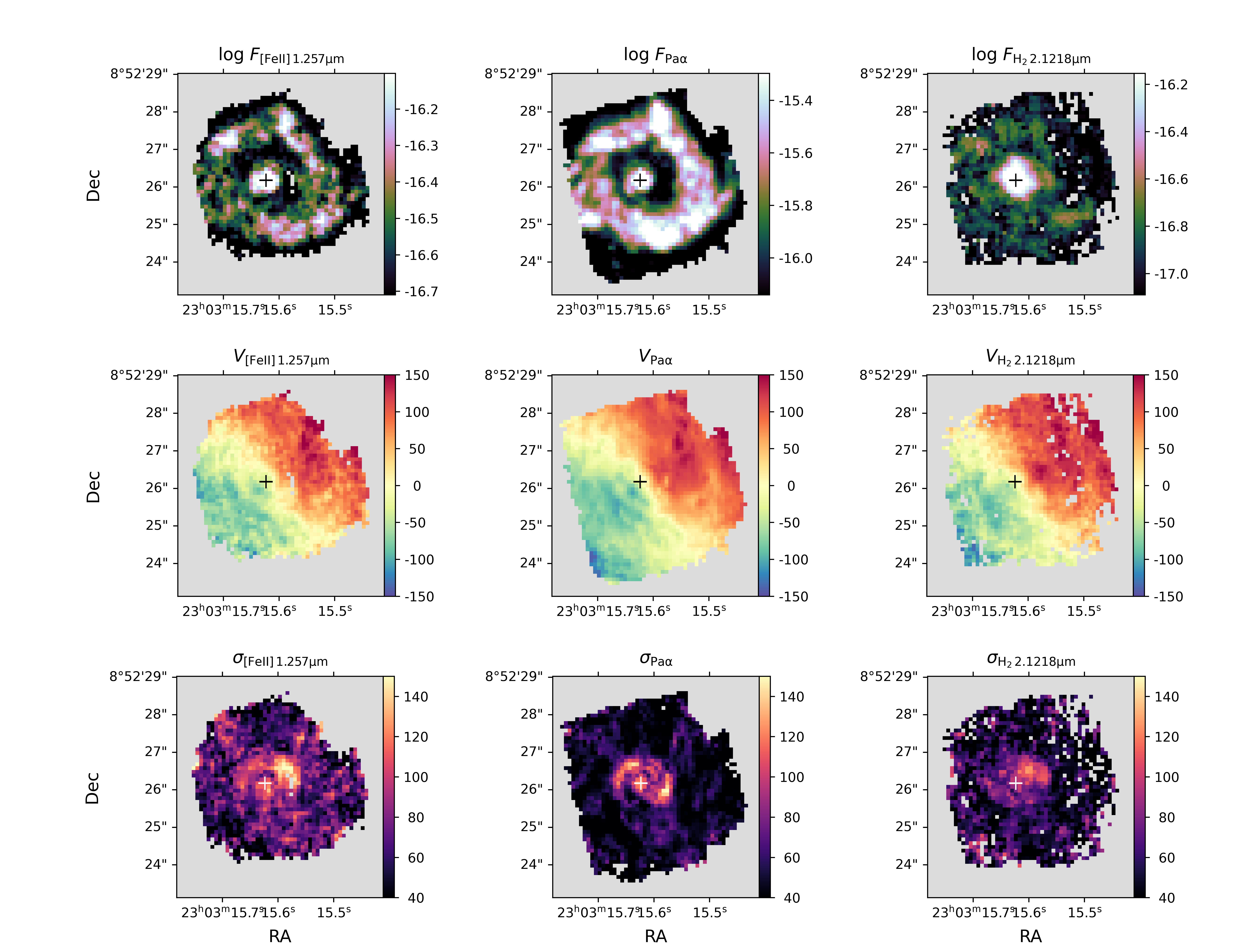}
    \caption{Flux (top row, in erg~s$^{-1}$~cm$^{-2}$, logarithmic scale), velocity (middle row, in km~s$^{-1}$) and velocity dispersion (bottom row, in km~s$^{-1}$) of three emission lines:  \feii\,1.257$\mu$m, \paa\ and \molhy\,2.1218$\mu$m. The cross indicates the position of the nucleus across all the panels. North is up and East to left. A connection  between the ring and the nucleus, to the southeast of the latter, is observed in the \paa~flux map. The gas motions are dominated by rotation, but non-circular motions also contribute to the gas dynamics as evidenced by the excesses of blueshift and redshifts to the southeast and northwest of the nucleus, respectively, and regions with enhanced velocity dispersion close to the nucleus, but with different morphologies for the three emission lines. The locations of the enhanced velocity dispersion are signatures of gas disturbed by the outflow, for \feii~and \molhy, or gas heated by friction in an inflow, for the \paa.  See Sec.\,\ref{sec:dynamics} for the discussion.
    }
    \label{fig:maps}
\end{figure*}

\section{Analysis and results}
\label{sec:results}
The right panels of Fig.\,\ref{fig:overview} show two spectra extracted from the nucleus ($0.4\arcsec$ radius, in black) and from the star-forming ring ($1\arcsec-1.8\arcsec$ radii, in blue) in NGC\,7469. In the nuclear spectrum, we detect prominent emission from the Hydrogen (H) recombination lines, particularly broad components in the Paschen series, and permitted lines such as O\,\textsc{i} and He\,\textsc{ii}. This broad component may be associated with the broad line region (BLR) of AGNs and is typical of type 1 sources. The nuclear spectrum also features coronal lines such as in \sivii~and \mgviii~with blueshifted profiles, similar to observations of higher ionization coronal lines from \citet{armus23}, indicative of gas outflows. Narrow lines such as \feii, \molhy~, and the PAH band at 3.3\um~are only weakly detected. The strong continuum radiation from the AGN might be the cause for the reduced equivalent width of these lines at the nucleus. On the other hand, PAH features and CO absorption bands are featured prominently in the ring spectrum, while coronal lines and the broad component of the H recombination lines disappear. We also see other ionized gas species such as [P \textsc{ii}] and [Fe \textsc{ii}], as well as various \molhy~transitions in the ring. 
{Tables \ref{tbl:f_g140}, \ref{tbl:f_g235}, \ref{tbl:f_g395} present the emission line integrated fluxes for each grating in both nuclear and ring extractions. The fluxes are obtained by fitting one, or two in the case of recombination and coronal lines, Gaussian functions to each line, and a polynomial to the continuum. Only the lines with at least a $3\sigma$ detection in each extraction are included. The fluxes of the 3.3$\mu$m PAH and aliphatic feature are part of our companion paper \cite{lai23}, which discusses the dust grain distribution in NGC\,7469 in detail.}

For the present paper, we focus our analysis on a subset of the emission lines observed (i.e. \feii, \molhy, and the H\,\textsc{i} lines) to highlight the conditions of the multiphase ISM. 
{
\begin{deluxetable}{lcccc}[htb]
\tablecaption{Emission Line Fluxes in the nucleus and SF ring in the G140H grating.\label{tbl:f_g140}}
\tablecolumns{8}
\tablewidth{0pt}
\tablehead{
\colhead{Line} &
\colhead{$\lambda_{rest}$}&
\colhead{IP} &
\colhead{Nucleus} &
\colhead{Ring}
}
\startdata
[C I] & 0.985 &   &  $4.56 \pm 0.75$ & $2.35 \pm 0.82$ \\ 
\pad & 1.005 &  & $77.18 \pm 8.31$ & $14.81 \pm 1.51$ \\
He II & 1.012& 24.6 & $22.9 \pm 7.4$ &  \\ 
He I &  1.083&  & $451.52 \pm 2.03$ & $57.86 \pm 3.54$ \\
Pa$\gamma$ & 1.094  &  & $122.67 \pm 2.02$ & $31.64 \pm 2.7$ \\
O I & 1.129 &  & $36.86 \pm 1.57$ & \\ 
$\rm\left[P~II\right]$ & 1.147& 10.5 & & $2.88 \pm 1.24$ \\ 
$\rm\left[P~II\right]$ & 1.188 & 10.5 & $2.44 \pm 0.07$ & $5.26 \pm 1.24$ \\ 
$\rm\left[S~IX\right]$ & 1.252& 328.8 & $4.46 \pm 0.39$ &\\ 
$\rm\left[Fe~II\right]$ & 1.257 & 7.9  & $6.62 \pm 0.24$ & $27.15 \pm 1.19$ \\ 
\pab & 1.281 &   & $153.6 \pm 1.84$ & $75.52 \pm 1.1$ \\ 
$\rm\left[Fe~II\right]$ & 1.321 & 7.9 & $3.24 \pm 0.36$ & $8.52 \pm 2.25$ \\ 
$\rm\left[Fe~II\right]$ & 1.372 & 7.9  & & $4.6 \pm 2.25$ \\ 
$\rm\left[Fe~II\right]$ & 1.644 & 7.9  & $6.76 \pm 1.37$ & $25.56 \pm 0.25$ \\ 
\enddata
\tablecomments{Rest wavelengths ($\lambda_{rest}$) are in units of $\mu$m, ionization potentials (IPs) in eV and all fluxes are in 10$^{-15}$\,erg\,s$^{-1}$\,cm$^{-2}$.
}
\end{deluxetable}
}

\begin{deluxetable}{lcccc}[htb]
\tablecaption{Emission Line Flux Densities in the nucleus and SF ring in the G235H grating.\label{tbl:f_g235}}
\tablecolumns{8}
\tablewidth{0pt}
\tablehead{
\colhead{Line} &
\colhead{$\lambda_{rest}$}&
\colhead{IP} &
\colhead{Nucleus} &
\colhead{Ring}
}
\startdata
Br10 & 1.7636 &   & &$6.94 \pm 1.39$ \\ 
H$_2$1-0S(5) & 1.835 &   & & $2.97 \pm 1.61$\\ 
Pa$\alpha$ & 1.875 &  & $224.31 \pm 0.23$& $246.57 \pm 1.28$\\ 
\brd &  1.945&   &&$9.84 \pm 1.28$\\ 
 H$_2$1-0S(3) & 1.957&   &$7.65 \pm 0.03$&$7.48 \pm 1.49$\\ 
  $\rm \left[Si~VI\right]$ & 1.963 & 166.8 &$14.35 \pm 0.06$ &\\ 
  H$_2$1-0S(2) & 2.033 &  &&$3.69 \pm 1.51$\\ 
  He I & 2.058 &   & $1.9 \pm 0.09$& $ 6.59 \pm 0.97$\\ 
   H$_2$1-0S(1) & 2.122&   &$7.39 \pm 0.41$&$11.23 \pm 1.4$\\ 
\brg & 2.166 &   &$30.44 \pm 2.67$&$24.51 \pm 1.21$\\
H$_2$1-0S(0) &  2.223&   &&$ 6.77 \pm 1.4$\\
H$_2$1-0Q(5) & 2.454 &   &$2.99 \pm 0.61$&\\
$\rm\left[Si~VII\right]$ &  2.483&  205.3& $17.93 \pm 0.96$&\\ 
\brb &  2.625&  &$36.99 \pm 6.84$& $44.29 \pm 1.85$\\
H$_2$1-0O(3) & 2.802 &   & &$7.75 \pm 2.15$\\
$\rm\left[Mg~VIII\right]$ &  3.027& 225.2 &$17.07 \pm 0.28$ &\\
Pf$\zeta$ & 3.039 & & &$3.68 \pm 1.25$\\
\enddata
\tablecomments{Rest wavelengths ($\lambda_{rest}$) are in units of $\mu$m, ionization potentials (IPs) in eV and all fluxes are in 10$^{-15}$\,erg\,s$^{-1}$\,cm$^{-2}$.
}
\end{deluxetable}

\begin{deluxetable}{lcccc}[htb]
\tablecaption{Emission Line Flux Densities in the nucleus and SF ring in the G395H grating.\label{tbl:f_g395}}
\tablecolumns{8}
\tablewidth{0pt}
\tablehead{
\colhead{Line} &
\colhead{$\lambda_{rest}$}&
\colhead{IP} &
\colhead{Nucleus} &
\colhead{Ring}
}
\startdata
H$_2$1-0O(5) & 3.235 &   && $5.3 \pm 0.59$ \\
Pf$\epsilon$ &  3.297&   && $5.17 \pm 1.68$\\
Pf$\delta$ & 3.740 &   &$12.36 \pm 2.98$& $10.56 \pm 0.2$\\
H$_2$0-0S(11) & 4.181 &  & $2.07 \pm 0.15$& $2.81 \pm 0.25$\\
Hu12 & 4.376 & & & $3.1 \pm 0.38$ \\
$\rm\left[Mg~IV\right]$ & 4.486 & 80.1 & $13.51 \pm 0.59$& $0.79 \pm 0.22$\\ 
$\rm\left[Ar~VI\right]$ & 4.529 &  74.8 &$8.97 \pm 0.57$ &\\ 
Pf$\gamma$ & 4.653 &  & $1.34 \pm 0.32$& $18.56 \pm 0.25$\\
Hu$\epsilon$ & 4.672 &   &&$2.59 \pm 0.25$\\
H$_2$0-0S(9) & 4.694 &   & $3.29 \pm 0.27$ & $5.38 \pm 0.25$\\
H$_2$0-0S(8) & 5.053 &   & $1.31 \pm 0.27$& $2.49 \pm 0.1$\\
Hu$\delta$ & 5.128 &  && $3.27 \pm 0.1$\\
\enddata
\tablecomments{Rest wavelengths ($\lambda_{rest}$) are in units of $\mu$m, ionization potentials (IPs) in eV and all fluxes are in 10$^{-15}$\,erg\,s$^{-1}$\,cm$^{-2}$.
}
\end{deluxetable}

Due to the under-sampling of the point spread function (PSF) in the NIRSpec IFS mode, the per-spaxel spectra in the nuclear region exhibit a wiggly pattern that is different from the fringes characteristic of 1d spectra extracted from MRS data. To mitigate this effect, we use a spectrum averaged over an aperture with a radius of $0.2\arcsec$ as representative of the nuclear region. The chosen aperture size is motivated by the size of the PSF. 

In order to obtain emission line moment maps, we use the Python package IFSCUBE \citep{ruschel-dutra20,ruschel-dutra21} to fit the emission lines of interest and with Gaussian functions and underlying continuum with a polynomial function in each spaxel of the data cube.  We fit a broad component {\citep[$\sigma\approx890$\,km\,s$^{-1}$ at systemic velocity, corresponding to $FWHM\approx2095$\,km\,s$^{-1}$ in agreement with][]{lu21}}typical of Type 1 AGNs, and a narrow component to the hydrogen recombination lines. {In the case of the Pa$\alpha$ line, an extra component was necessary to reproduce the broad ($\sigma\approx 800$\,km\,s$^{-1}$) emission line observed beyond the PSF extension, likely associated with a gas outflow.} 
A single narrow component is enough to reproduce the emission line profiles of the forbidden and molecular hydrogen lines.
A third-degree polynomial reproduced the underlying continuum at the gratings G140H and G235H. The fitting in each grating is performed independently, but within each grating, we tie the systemic velocity and velocity dispersion of the lines that trace the same gas phase.

\subsection{Flux and Line Ratios}
The resulting moment maps from our spectral fitting procedure for three emission lines representative of the atomic ionized --- \feii~$1.257\mu$m and \paa\ --- and the molecular gas --- \molhy~$2.12\mu$m --- phases are shown in  Figure\,\ref{fig:maps}. 
Both \feii~and \paa~have low ionization potentials, with IP = 7.9~eV and 13.6~eV, respectively, and the integrated flux maps show that they trace the star-forming ring and the point source at the nucleus. The ring has knots of star formation to the North, West, and South, but is fainter and more flocculent to the East of the nucleus. 
The 0\farcs2~resolution from NIRSpec, at 1.9$\mu$m, reveals a tail southeast of the nucleus in the inner ISM region that, in projection, connects the ring to the AGN as seen in the \paa~flux map {(see Fig.\,\ref{fig:maps})}. 
The inner ISM region has surface brightness around an order of magnitude lower than the ring and the nucleus, indicating that the excitation/heating of the gas might be different in this region. The \molhy~flux is more concentrated at the center where the peak of the emission is also observed. It does not show a ring-like pattern; instead, it displays a more extended and diffuse morphology, resembling a spiral arm to the Southwest as previously noticed in the warm molecular phase with MIRI \citep{u22c} and cool CO gas \citep{davies04}. Both \molhy\,$2.12\mu$m and \molhy\,$1.95\mu$m fluxes have this distribution. 

\begin{figure*}
    \centering
\includegraphics[trim={1.5cm 1.5cm 1.5cm 1.5cm},clip, width=0.9\textwidth]{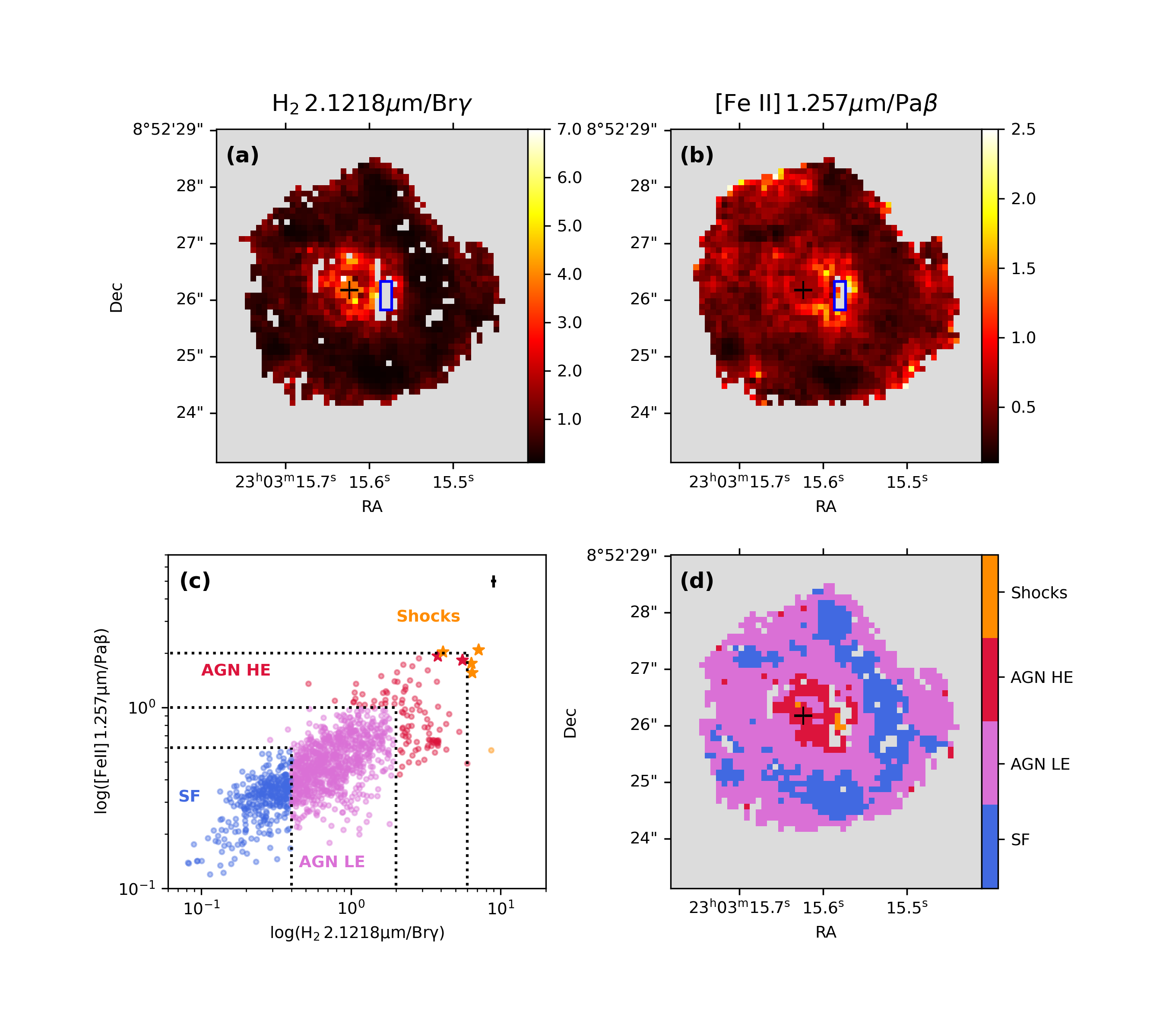}
    \caption{Line ratio diagnostics in the near-IR: \molhy~2.1218$\mu$m/\brg~(a) and \feii~1.257$\mu$m/\pab~(b). Given that the \pab~and \brg~lines in the inner-ISM region are only marginally detected (SNR$<2$), lower limits of the \molhy/\brg~and \feii/\pab~ratios are estimated using the values inside the blue boxes and presented as the star symbols in the near-IR excitation diagnostic diagram (c). 
    The bottom panels show the near-IR excitation diagram \cite[(c); adapted from][]{riffel13} and the corresponding spatial location of spaxels dominated by different excitation (d). 
    The spaxels are color-coded by their location on the excitation diagram: star formation (SF), AGN low excitation (LE), AGN high excitation (HE), and shocks. The line ratios encompass the complexity of the gas excitation in the inner kiloparsec of NGC\,7469, where evidence of star formation and, the AGN radiation field and high excitation due to shocks are present. {The typical uncertainties in both line ratios are indicated in the top right corner of panel (c).}
    }
    \label{fig:excitation}
\end{figure*}

In order to investigate the origin of the gas emitted in the central region of NGC\,7469, we create emission line ratio maps of \molhy/\brg~and \feii~$1.257\mu$m/\pab~(Fig.~\ref{fig:excitation}, panels (a) and (b)) which are insensitive to extinction. The ring shows low line ratio values of $\sim0.5$ for both \molhy/\brg~and \feii~$1.257\mu$m/\pab. The \molhy/\brg~has the highest values ($\sim 3.5$) at the nucleus and in the western inner ISM region. The \feii/\pab~ is also the highest in the latter. The gray regions correspond to spaxels where at least one of the two lines in the ratio has a signal-to-noise ratio (SNR) lower than 2. In the inner-ISM region, the low SNR spaxels are attributed to \pab~and \brg~non-detections.

The excitation diagram (Figure \ref{fig:excitation} (c)) shows data points based on the emission line ratios presented in panels (a) and (b).
To distinguish among the different mechanisms that regulate these line ratios, we use the thresholds derived by \citet{riffel13}, and
color-code the spaxels depending on their location on the excitation diagram (panel (d)). 
These boundaries are derived from 
analyzing single-slit spectra of galaxies previously classified as starbursts, AGN hosts, or shock-dominated sources such as LINERS. 
The division between AGN low excitation (LE) and AGN high excitation (HE) is motivated by allowing for a finer sampling of the AGN influence on spatial scales. 
We use \feii~$1.257\mu$m / \pab~$=1$ and \molhy~$2.12\mu$m / \brg~$=2$ as adopted by \citet{riffel21_llp_exc} which are arbitrary. 

Since the recombination lines have low fluxes at the inner-ISM region and the \molhy~is detected at the same region, we expect high line ratios in the western part of the nucleus. The blue box overlaid in the line ratio maps indicates the region where we extracted average spectra for the G140H and G235H gratings. From these spectra, we obtain upper limits for the \brg~and \pab~emission lines. 
The ratios are then obtained using the flux per spaxel from our \molhy~and \feii~maps and are plotted on the excitation diagram as stars indicating they are lower limits on the \molhy~2.12$\mu$m/\brg~ ratio.

Applying the color-coding from the excitation diagram to the excitation map (Figure~\ref{fig:excitation} panel (d)), we find that the spaxels with the largest flux ratios are consistent with photoionization by the AGN. A small number of spaxels may be shock excited. 
The inner ISM region has line ratios consistent excitation by the AGN radiation field and by shocks. With the exception of the starforming ring itself, the bulk of the ISM probed by the NIRSpec IFS is photoionized by the AGN.

\subsection{Velocity and velocity dispersion}

In the \feii\,$1.257\mu$m, \paa~and \molhy\,$2.1218\mu$m velocity fields (middle row of Figure~\ref{fig:maps}), rotation is the dominant feature, with amplitudes of up to 150\,km\,s$^{-1}$. The rotating disk of hot \molhy~we measure was first detected with adaptive optics
spectroscopy on the Keck telescope \citep{hicks08}.
That study found similar rotation velocities to what we see: 100\,km\,s$^{-1}$ or more
out to a radius of about $0.8\arcsec$, almost reaching the starburst ring.
Small-scale kinematic features are also observed as excess blueshifts to the southeast and redshifts to the northwest of the nucleus. The velocity field for \pab~(not presented here) shows very similar kinematics and gas distribution as those of \paa, although with a lower SNR. 

The velocity dispersion maps show some interesting features previously seen at lower resolution in the MIRI-MRS data \citep{u22c}. The ring and outer parts have low dispersion values ($\sim 60-80$\,km\,s$^{-1}$) indicating that they are dominated by rotation. In the inner ISM region, we see differences between the gas species. The \feii~velocity dispersion map shows an arc-shaped structure with the highest values ($\sigma\approx140$\,km\,s$^{-1}$) to the west of the nucleus. 
For the \paa~we also see arc-shaped structures peaking in two different regions, to the northeast and southwest of the nucleus, with a dual-spiral pattern. 
Each of the spirals is located $\sim 0.6\arcsec$ ($\approx200$\,pc) away from the nucleus. We investigate the nature of this feature in Section\,\ref{sec:dynamics}.
The \molhy~velocity dispersion map displays higher values ($\sigma\approx120$\,km\,s$^{-1}$) extending 
 $0.9\arcsec$ northwest of the nucleus. This feature is also seen in the \molhy\,$1.95\mu$m velocity dispersion map (not presented here). The fan-shaped feature is consistent with the regions of enhanced \molhy~dispersion seen with MIRI-MRS \citep{u22c} but at a higher spatial resolution.


\begin{figure*}
    \centering
    \includegraphics[width=0.9\textwidth]{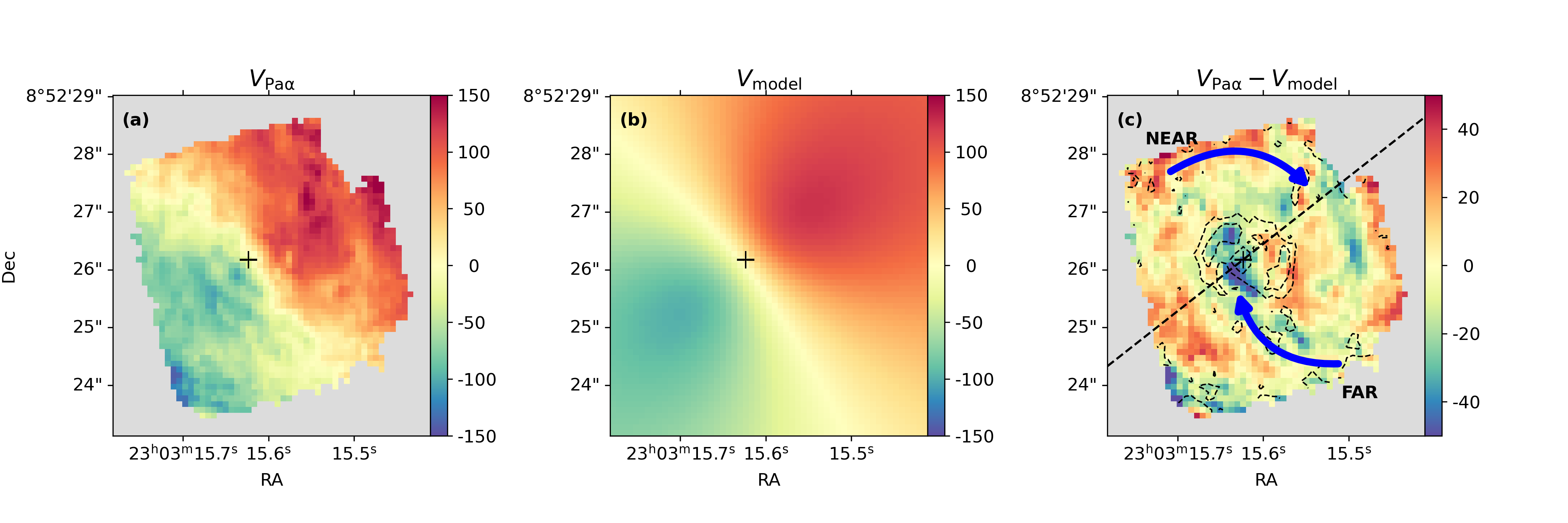}
    \caption{(Left to right) Observed \paa~velocity field, modeled velocity field with rotation described by a thin disk model, and the residual map from the difference between the observed and modeled velocity fields. The dashed line indicates the direction of the galaxy's kinematic major axis. The Near and Far sides of the disk are labeled, and the arrows indicate the location of gas that does not follow the galactic rotation and may be inflowing to the nucleus. The contours show the regions of increased velocity dispersion in \paa.}
    \label{fig:vel_model}
\end{figure*}

One way of investigating the non-rotational motions is by fitting the velocity fields with a rotating disk model that assumes the gas particles have circular orbits.
We adopt the model described by the following equation~\citep{bertola91}:
\begin{align*}
    &V=V_{\rm sys}+\\
    &\frac{AR\cos(\psi-\psi_0)\sin\theta\cos^{p}\theta}{\{R^2[\sin^2{(\psi-\psi_0)}+\cos^2(\theta)\cos^2{(\psi-\psi_0)}]+C_0\cos^2{\theta}\}^{p/2}},
    \label{eq:vel} 
\end{align*}

\noindent where $V_{sys}$ is the systemic velocity of the galaxy, $R$ is the distance of each pixel to the center of rotation, $A$ is the velocity amplitude, $\psi$ is the position angle of each spaxel, $\psi_0$ is the PA of the line of the nodes, i.e. the kinematical major axis, and $\theta$ is the inclination of the disk. The parameter $p$ is the slope of the rotation curve, varying from 1 for an asymptotically flat rotation curve to 1.5 for a system with finite mass, and $C_0$ is the concentration parameter, the radius where the velocity reaches 70\% of the velocity amplitude. 

We fit this model to the \paa~velocity field as this emission line has the most interesting structure. The fitting is performed using the non-linear least-square minimization routine \texttt{mpfit2dfun} \citep{markwardt09}. The combination of parameters that best describe the velocity field are $A=476.9\pm6.09$\,km\,s$^{-1}$, $V_{sys}=13.29\pm0.41$\,km\,s$^{-1}$ and $C_0=0.53\pm0.01$ arcsec, $\psi_0=128.81^{\circ}\pm0.38$ and $\theta=15.97^{\circ}\pm0.28$. The position of the center of rotation is adopted as the position of the nucleus, and $p=1.5$ are fixed during the fitting. Previous studies have estimated similar values for $\Psi_0$ \citep{davies04,hicks08, u22c} and $\theta$ \citep{hicks08}. 

Figure~\ref{fig:vel_model} shows the modeled velocity field in panel (b), and the residual velocity map, with the position angle of the galaxy's major kinematic axis marked by the black dashed line, in panel (c). The Near and Far side indicate the orientation of the star-forming ring proposed by \citet{u22c} and by the optical large-scale image of the galaxy. We observe most of the velocity residuals as blueshifts on the far side of the galaxy. Such a configuration may represent streaming motions toward the center.

\section{Discussion}
\label{sec:discussion}
\subsection{Gas excitation}

Emission line diagnostic diagrams are powerful tools to investigate the origin of emission lines in galaxies. In the near-IR, empirical diagrams involving bright lines in the $J$ and $K$ bands have been analyzed in single-aperture \citep{larkin98,reunanen02, alberto04, alberto05, riffel13} and spatially resolved studies \citep{colina15,u19,riffel21}. The \molhy~can be excited by soft-UV photon absorption (non-thermal) or by collisional excitation (thermal). The thermal processes include shocks \citep{hollenbach89, riffel15}, and X-ray heating by the AGN radiation field. The \feii~is produced in the partially ionized region that, in AGN hosts, is a byproduct of X-rays or shocks due to the radio jet or gas outflows \citep{simpson96,forbes93}. Another source of excitation of the \feii~is due to shocks caused by supernovae remnants, especially in star-forming regions \citep{rosenberg12}. 
Previous studies show that there is a correlation between \molhy/\brg~ and \feii$1.257\mu$m/\pab, suggesting that both \feii~ and \molhy~may have a common excitation mechanism \citep{larkin98,riffel13,riffel21}. A positive trend is observed in Figure \ref{fig:excitation} (c), suggesting that in NGC\,7469 \feii~and \molhy~are excited by a common mechanism. 


The excitation map (Figure \ref{fig:excitation} (d)) shows the spatial distribution of the different line ratios. 
At the location of the ring, the excitation conditions show a discontinuity to the east, where AGN-like excitation dominates. A morphological discontinuity in the ring, at the same location,  has previously been claimed as the result of an outflow-excavated ring \citep{garcia-bernete22}. However, we do not see clear velocity signatures consistent with outflowing gas in the spectral transitions we analyze here. 

The highest line ratios (color-coded as red and orange) are distributed mostly in the western part of the inner-ISM. These line ratios are indicative of shock-excited gas at the same location as in the MIRI-MRS observations of NGC\,7469 \citep{u22c}. The authors linked the shock-excited \molhy~with the fan-shaped \molhy~S(5) velocity dispersion map. The lower right panel in Fig.~\ref{fig:maps} shows a more clumpy structure than in \molhy~S(5). 
Shock models suggest that regions with fast motions would be expected in a clumpy turbulent medium \citep{appleton23}. Thus, the high-velocity dispersion and the shock-like line ratios are indicators of the same shocked gas structure at the western part of the inner-ISM. 

NGC\,7469 has a compact radio jet-core structure that extends east-west for $\sim$100\,pc, which  has been resolved with VLBI observations \citep{lonsdale03,alberdi06} and high-resolution VLA observations \citep{orienti10,song22} at multiple frequencies. \feii~is likely excited by shocks due to the interaction of the jet with the ISM in the regions color-coded in orange in Figure \ref{fig:excitation}. However, excitation by the AGN radiation field cannot be ruled out, as the jet is compact and line ratios consistent with excitation by the AGN are observed over all the FoV. 
The ring does show line ratios typical of star-forming galaxies, where non-thermal processes (UV-fluorescence) might be the mechanism responsible for the \molhy~excitation. The \feii~emission observed in the SF ring is likely due to shocks caused by supernovae where the dust grains trapping the Fe are destroyed by the supernova remnant shocks \citep{rosenberg12}.

\subsection{Gas dynamics: inflows and the feedback driving mechanism}
\label{sec:dynamics}
In Figure \ref{fig:maps}, the velocity dispersion of \paa~shows a dual spiral pattern, which also appears to exhibit enhanced \molhy/\brg~and \feii/\pab~ratios.
With optical observations of the AGN host Mrk\,590, \citet{raimundo19} observe a nuclear dual spiral pattern in the velocity dispersion and flux maps of optical ionized gas emission lines. The authors interpret this pattern as due to a gas inflow to the nucleus of the galaxy. Another example of a nuclear dual spiral is interpreted as a gas inflow in NGC\,6951 \citep{thaisa07}. In the case of NGC\,7469, the hypothesis of an inner dual spiral was proposed by \citet{davies04} to explain the morphology of the CO(2-1) flux distribution. However, the authors did not find kinematic signatures that could be associated with a gas inflow. In magneto-hydrodynamical simulations, stellar feedback from a star-forming ring provides the fuel for gas inflows that star and sustain the AGN-duty cycle \citep{clavijo23}, which is probably the case in NGC\,7469. We investigate possible deviations from pure circular rotation by inspecting the velocity residuals in Figure\,\ref{fig:vel_model} (c).
{The resolving power of NIRSpec IFU at the wavelength of the Pa$\alpha$ line is $R\approx2200$ \citep{jakobsen22}, which corresponds to a velocity resolution of $\approx130$\,km\,s$^{-1}$. the residuals are more prominent closer to the nuclear spiral, reaching $64$\,km\,s$^{-1}$ in blueshifts and patchy and with lower amplitudes in the outer regions of the field. The localized nature of the highest residuals in arm-shaped morphology leads us to interpret them as a signature of gas inflow. Similar approaches have been adopted in the literature previously \citep[e.g.][]{schnorr-muller16,schnorr-muller17, riffelugc}.}

If we indeed are observing inspiraling gas, we follow \citet{thaisa07} and compute the mass inflow rate via 
\begin{equation}
\dot{M}_{\rm in} = N_e V_{\rm in} A m_p f n_{\rm arms} \quad ,
\end{equation}

\noindent where $N_e$ is the electron density, $V_{\rm in}$ is the velocity of the inflowing gas, $A$ is the area of the cross-section of the spiral arm, $m_p$ is proton mass, $f$ is the filling factor and $n_{\rm arms}$ is the number of arms (in our case:  $n_{\rm arms}=2$). 
 We estimate the filling factor considering that 
 
\begin{equation}
    L_{\rm Pa\alpha}=f j_{\rm Pa\alpha}(T)v \quad , 
\end{equation}

\noindent where $L_{\rm Pa\alpha}$ and  $j_{Pa\alpha}(T)$ are the luminosity and emission coefficient of the \paa~line \citep{osterbrock}, and $v$ is the volume of the emitting region. Assuming a typical density of 500\,cm$^{-3}$ and a temperature of 10,000\,K, we determine $j_{\rm Pa\alpha}=8.29\times10^{-22}$ erg\,cm$^{-3}$\,s$^{-1}$. The \paa~luminoisity is estimated from the flux map in Fig.~\ref{fig:maps} as $2.4\times10^{41}$\,erg\,s$^{-1}$. Since most of the \paa~emission is associated with the star-forming ring, we assume that half of the line flux comes from the gas inflowing  to center as $L_{\rm Pa\alpha}=1.2\times10^{41}$\,erg\,s$^{-1}$. Assuming that the inflowing gas is distributed in a dual-spiral structure where  one of the arms can be represented by a cylinder of height $h=1.5\arcsec$ and radius $r=0.3\arcsec$, the filling factor is $f=0.14$. This value is higher than previously estimated in the nearby Seyfert galaxy NGC\,6951 $f=0.004$ \citep{thaisa07}. The mass inflow calculation adopts both filling factor values.  The velocity is corrected by the inclination of the galaxy ($V_{\rm in}=\frac{V_{\rm res}}{\sin{i}}$). From the rotation model residuals (Fig.\,\ref{fig:vel_model}, panel (c)) $V_{\rm res}\approx40$km\,s$^{-1}$. We consider two possible inclinations for NGC\,7469: $i=16^{\circ}$ derived from the rotation model, and $i=51^{\circ}$ the inclination of the ring \citep{u22c}. The deprojected inflow velocity is  $V_{\rm in}=51-145$\,km\,s$^{-1}$.

Considering all the assumptions mentioned above, the mass inflow rate is $\dot{M}_{\rm in}=0.2-17$\,M$_{\odot}$\,yr$^{-1}$. This value is comparable to the mass outflow rate estimated from the coronal emission lines \citep[$1-5$\,M$_{\odot}$yr$^{-1}$]{muller-sanchez11,armus23}. The mass inflow rate is at least one order of magnitude larger than the accretion rate needed to power the AGN \cite[$\sim4.1\times10^{-2}$\,M$_{\odot}$yr$^{-1}$;][]{armus23}. Part of the gas mass moving towards the center might not feed to the supermassive black hole and could, in fact, be ejected as part of the outflow. 

{The fitting of the brightest hydrogen recombination line observed in NGC\,7469 NIRSpec IFU data reveals the presence of a compact ($R=0.6\arcsec\equiv205$\,pc), high dispersion ($\sigma=800$\,km\,s$^{-1}$) at systemic velocity ($V=11$\,km\,s$^{-1}$) component. We interpret this component as an outflow, as is expected from the presence of gas outflows in highly \citep{muller-sanchez11,armus23} and moderately \citep{xu22} ionized gas in this galaxy. In order to investigate the impact of this outflow in the galaxy, we proceed to calculate the mass outflow rates following two different methods. The first follows equation 4 in \citet{bianchin22}  \citep[see also][]{harrison14, fiore17, kakkad20,riffelagnifskin}. The ionized outflow gas mass ($M_{\rm out}\approx10^6$\,M$_{\odot}$) is obtained by adapting the equation 5 from \citet{storchi-bergmann09} where $F_{\rm Pa\alpha}/F_{\rm Br\gamma}=12.19$, assuming the case B of recombination B recombination, a temperature of 10,000\,K and an electron density of 500\,cm$^{-3}$. The resulting mass outflow rate is $\dot{M}_{\rm out}=0.18$\,M$_{\odot}$yr$^{-1}$. The second method follows \citet{muller-sanchez11,armus23}, we use the same electron density as in the first method, and a filling factor $f=0.001$. Instead of the biconical geometry adopted by \citet{armus23}, we adopt a spherical geometry due to the lack of evidence of a biconical structure in the outflow. The resulting mass outflow following this method is $\dot{M}_{\rm out}=0.14$\,M$_{\odot}$yr$^{-1}$, in good agreement with the previous method. This mass outflow is consistent with the values derived from the analysis of the outflows traced by the Pa$\beta$ emission line in local Seyferts \citep{bianchin22}}. 

{The mechanical impact of the outflow in the ISM can be accessed via its kinetic power determined by $\dot{E}_{out}=\dot{M}_{\rm out}(V_{\rm max}^2+3\sigma^2)/2$. The $\sigma^2$ term is particularly important for this case, as most of the energy is associated with turbulent motions. Adopting the velocity, velocity dispersion, and mass outflow rates determined previously, $\dot{E}_{out}= 5.9-7.2\times10^{36}$\,erg\,s$^{-1}$. }
In order to determine if the small-scale radio jet-core structure is capable of driving the highly ionized gas outflow, we follow \citet{morganti15} and \citet{venturi21} and adopted the scaling relations between 1.4\,GHz radio luminosity and jet power presented in \citet{birzan08} and \citet{cavagnolo10}. 
The jet component has flux density of 9.4\,mJy at 8\,GHz \citep{alberdi06}, and 1.3\,mJy at 33\,GHz \citep{song22}, which yield a $8 - 33$\,GHz spectral index of $\alpha \sim -1.4$ and an extrapolated 1.4 GHz flux density of $\sim$\,120\,mJy. This value would be lowered to 40\,mJy if a nominal synchrotron spectral index of $\alpha_{\rm NT} = -0.8$ is assumed between 1.4 and 8\,GHz. Using the scaling relations, these values correspond to kinetic jet power of $0.6-3\times10^{43}$\,erg\,s$^{-1}$. As noted in \citet{venturi21}, jet power may be underestimated by an order of magnitude using the above scaling relations compared to values derived from models of jet-ISM interaction \citep[e.g.][]{mukherjee18}. The kinetic power of the outflow is {several orders} of magnitude smaller than the jet power {implying that the jet can perturb the gas in its vicinity, increasing its turbulence and thus being the main of the low velocity and high dispersion outflow observed in the Pa$\alpha$ emission line.}

\section{Summary}
\label{sec:conclusions}
{In this paper}, we report the {\em JWST} NIRSpec-IFS observations of NGC\,7469. With the superb spatial and spectral resolutions of NIRSpec, it is now possible to study in detail the inner kiloparsec of this galaxy. This allows us to unveil the intricate connection between the feeding, feedback, and gas excitation in the inner kiloparsec of this luminous Sy\,1.5 galaxy. 
Our conclusions are summarized below. 

\begin{itemize}
    \item Enhanced line ratios, consistent with shock-heated gas, are observed mostly in the nuclear region of the galaxy, with the highest line ratios to the west of the nucleus. 
    The presence of a compact ($<100$\,pc) radio jet \citep[e.g.][]{lonsdale03} indicates that its interaction with the gas in the inner-ISM region may be the cause of the \molhy~and \feii~gas excitation. The spatially resolved excitation map also reveals diffuse AGN-like excitation over the field of view suggesting that the central source ionizes a significant fraction of the gas in the inner ISM.

    \item NGC\,7469 kinematics is dominated by rotation (with amplitudes $\sim 150$\,km\,s$^{-1}$ at low-velocity dispersion ($60-80$\,km~s$^{-1}$), especially in the star-forming ring region.  We identify a nuclear spiral in NGC\,7469 in the \paa~velocity dispersion map. 
    We interpret the nuclear spiral and the non-rotational motions (observed after modeling the rotation field) as a gas inflow to the nucleus of the galaxy. We estimate a mass inflow rate of $0.2-17$\,M$_{\odot}$yr$^{-1}$ dependent upon the precise geometry and inclination of the disk. This value is up to two orders of magnitude higher than the gas accretion rate needed to power the central AGN but comparable with the mass outflow rate. 

    \item {The Pa$\alpha$ emission line has a kinematical component consistent with an ionized gas outflow. This outflow is dominated by turbulent motions. It carries a kinetic power of $5.9-7.2\times10^{36}$\,erg\,s$^{-1}$ which is consistent with being driven by the compact nuclear radio jet with a power of $10^{43}$\,erg\,s$^{-1}$.}
    
\end{itemize}

\begin{acknowledgments}
We thank Sandra Raimundo and Rogemar A. Riffel for the discussion that added to the interpretations in the paper. This work is based on observations made with the NASA/ESA/CSA James Webb Space Telescope. The data were obtained from the Mikulski Archive for Space Telescopes at the Space Telescope Science Institute, which is operated by the Association of Universities for Research in Astronomy, Inc., under NASA contract NAS 5-03127 for JWST. These observations are associated with program \#JWST-ERS-01328 and can be accessed via \dataset[DOI: 10.17909/w7ea-wx90]{https://doi.org/10.17909/w7ea-wx90}. 
Research at UCI by M.B. and V.U was supported by funding from program \#JWST-GO-01717, which was provided by NASA through a grant from the Space Telescope Science Institute, which is operated by the Association of Universities for Research in Astronomy, Inc., under NASA contract NAS 5-03127.
V.U further acknowledges partial funding support from NASA Astrophysics Data Analysis Program (ADAP) grants \#80NSSC20K0450 and \#80NSSC23K0750, and HST grants \#HST-AR-17063.005-A and \#HST-GO-17285.001, and had partially performed work for this project at the Aspen Center for Physics, which is supported by National Science Foundation grant PHY-2210452. 
The Flatiron Institute is supported by the Simons Foundation. H.I. and T.B. acknowledge support from
JSPS KAKENHI grant No. JP21H01129 and the Ito Foundation for Promotion of Science. A.M.M. acknowledges support from the National Science Foundation under grant No. 2009416. A.S.E. and S.L. acknowledge support from NASA grant HST-GO15472. Y.S. was funded in part by the NSF through the Grote Reber Fellowship Program administered by Associated Universities, Inc./National Radio Astronomy Observatory. The National Radio Astronomy Observatory is a facility of the National Science Foundation operated under cooperative agreement by Associated Universities, Inc. S.A. gratefully acknowledges support from an ERC Advanced grant 789410, from the Swedish Research Council and from the Knut and Alice Wallenberg (KAW) Foundation. 
S.T.L was partially supported thorough NASA grant HST-GO16914.
K.I. acknowledges support by the Spanish MCIN under grant PID2019 105510GBC33/AEI/10.13039/501100011033. 
F.M-S. acknowledges support from NASA through ADAP award 80NSSC19K1096. 
This work was also partly supported by the Spanish program Unidad de Excelencia María de Maeztu CEX2020-001058-M, financed by MCIN/AEI/10.13039/501100011033.
Finally, this research has made use of the NASA/IPAC Extragalactic Database (NED), which is operated by the Jet Propulsion Laboratory, California Institute of Technology, under
contract with the National Aeronautics and Space Administration.

\end{acknowledgments}
%

\newpage
\vspace{5mm}
\facilities{JWST (NIRSpec IFU and NIRCam), HST (WFC3), MAST, NED.}

\vspace{10mm}
\software{Astropy \citep{astropy:2013, astropy:2018, astropy:2022},  
IFSCUBE \citep{ruschel-dutra20,ruschel-dutra21},
{\em JWST} Science Calibration \citep{bushouse2022}, 
Matplotlib \citep{hunter07}, 
QFitsView \citep{ott12},
SciPy \citep{SciPy}.
          }

\bibliography{sample631}{}

\begin{thebibliography}{}
\expandafter\ifx\csname natexlab\endcsname\relax\def\natexlab#1{#1}\fi
\providecommand{\url}[1]{\href{#1}{#1}}
\providecommand{\dodoi}[1]{doi:~\href{http://doi.org/#1}{\nolinkurl{#1}}}
\providecommand{\doeprint}[1]{\href{http://ascl.net/#1}{\nolinkurl{http://ascl.net/#1}}}
\providecommand{\doarXiv}[1]{\href{https://arxiv.org/abs/#1}{\nolinkurl{https://arxiv.org/abs/#1}}}

\bibitem[{{Aalto} {et~al.}(2019){Aalto}, {Muller}, {K{\"o}nig}, {Falstad},
  {Mangum}, {Sakamoto}, {Privon}, {Gallagher}, {Combes}, {Garc{\'\i}a-Burillo},
  {Mart{\'\i}n}, {Viti}, {van der Werf}, {Evans}, {Black}, {Varenius},
  {Beswick}, {Fuller}, {Henkel}, {Kohno}, {Alatalo}, \& {M{\"u}hle}}]{aalto19}
{Aalto}, S., {Muller}, S., {K{\"o}nig}, S., {et~al.} 2019, \aap, 627, A147,
  \dodoi{10.1051/0004-6361/201935480}

\bibitem[{{Alberdi} {et~al.}(2006){Alberdi}, {Colina}, {Torrelles}, {Panagia},
  {Wilson}, \& {Garrington}}]{alberdi06}
{Alberdi}, A., {Colina}, L., {Torrelles}, J.~M., {et~al.} 2006, \apj, 638, 938,
  \dodoi{10.1086/498859}

\bibitem[{{Appleton} {et~al.}(2023){Appleton}, {Guillard}, {Emonts},
  {Boulanger}, {Togi}, {Reach}, {Alatalo}, {Cluver}, {Diaz Santos}, {Duc},
  {Gallagher}, {Ogle}, {O'Sullivan}, {Voggel}, \& {Xu}}]{appleton23}
{Appleton}, P.~N., {Guillard}, P., {Emonts}, B., {et~al.} 2023, arXiv e-prints,
  arXiv:2301.02928, \dodoi{10.48550/arXiv.2301.02928}

\bibitem[{{Armus} {et~al.}(2023){Armus}, {Lai}, {U}, {Larson}, {Diaz-Santos},
  {Evans}, {Malkan}, {Rich}, {Medling}, {Law}, {Inami}, {Muller-Sanchez},
  {Charmandaris}, {van der Werf}, {Stierwalt}, {Linden}, {Privon},
  {Barcos-Mu{\~n}oz}, {Hayward}, {Song}, {Appleton}, {Aalto}, {Bohn},
  {B{\"o}ker}, {Brown}, {Finnerty}, {Howell}, {Iwasawa}, {Kemper}, {Marshall},
  {Mazzarella}, {McKinney}, {Murphy}, {Sanders}, \& {Surace}}]{armus23}
{Armus}, L., {Lai}, T., {U}, V., {et~al.} 2023, \apjl, 942, L37,
  \dodoi{10.3847/2041-8213/acac66}

\bibitem[{{Arribas} {et~al.}(2014){Arribas}, {Colina}, {Bellocchi}, {Maiolino},
  \& {Villar-Mart{\'\i}n}}]{arribas14}
{Arribas}, S., {Colina}, L., {Bellocchi}, E., {Maiolino}, R., \&
  {Villar-Mart{\'\i}n}, M. 2014, \aap, 568, A14,
  \dodoi{10.1051/0004-6361/201323324}

\bibitem[{{Astropy Collaboration} {et~al.}(2013){Astropy Collaboration},
  {Robitaille}, {Tollerud}, {Greenfield}, {Droettboom}, {Bray}, {Aldcroft},
  {Davis}, {Ginsburg}, {Price-Whelan}, {Kerzendorf}, {Conley}, {Crighton},
  {Barbary}, {Muna}, {Ferguson}, {Grollier}, {Parikh}, {Nair}, {Unther},
  {Deil}, {Woillez}, {Conseil}, {Kramer}, {Turner}, {Singer}, {Fox}, {Weaver},
  {Zabalza}, {Edwards}, {Azalee Bostroem}, {Burke}, {Casey}, {Crawford},
  {Dencheva}, {Ely}, {Jenness}, {Labrie}, {Lim}, {Pierfederici}, {Pontzen},
  {Ptak}, {Refsdal}, {Servillat}, \& {Streicher}}]{astropy:2013}
{Astropy Collaboration}, {Robitaille}, T.~P., {Tollerud}, E.~J., {et~al.} 2013,
  \aap, 558, A33, \dodoi{10.1051/0004-6361/201322068}

\bibitem[{{Astropy Collaboration} {et~al.}(2018){Astropy Collaboration},
  {Price-Whelan}, {Sip{\H{o}}cz}, {G{\"u}nther}, {Lim}, {Crawford}, {Conseil},
  {Shupe}, {Craig}, {Dencheva}, {Ginsburg}, {Vand erPlas}, {Bradley},
  {P{\'e}rez-Su{\'a}rez}, {de Val-Borro}, {Aldcroft}, {Cruz}, {Robitaille},
  {Tollerud}, {Ardelean}, {Babej}, {Bach}, {Bachetti}, {Bakanov}, {Bamford},
  {Barentsen}, {Barmby}, {Baumbach}, {Berry}, {Biscani}, {Boquien}, {Bostroem},
  {Bouma}, {Brammer}, {Bray}, {Breytenbach}, {Buddelmeijer}, {Burke},
  {Calderone}, {Cano Rodr{\'\i}guez}, {Cara}, {Cardoso}, {Cheedella}, {Copin},
  {Corrales}, {Crichton}, {D'Avella}, {Deil}, {Depagne}, {Dietrich}, {Donath},
  {Droettboom}, {Earl}, {Erben}, {Fabbro}, {Ferreira}, {Finethy}, {Fox},
  {Garrison}, {Gibbons}, {Goldstein}, {Gommers}, {Greco}, {Greenfield},
  {Groener}, {Grollier}, {Hagen}, {Hirst}, {Homeier}, {Horton}, {Hosseinzadeh},
  {Hu}, {Hunkeler}, {Ivezi{\'c}}, {Jain}, {Jenness}, {Kanarek}, {Kendrew},
  {Kern}, {Kerzendorf}, {Khvalko}, {King}, {Kirkby}, {Kulkarni}, {Kumar},
  {Lee}, {Lenz}, {Littlefair}, {Ma}, {Macleod}, {Mastropietro}, {McCully},
  {Montagnac}, {Morris}, {Mueller}, {Mumford}, {Muna}, {Murphy}, {Nelson},
  {Nguyen}, {Ninan}, {N{\"o}the}, {Ogaz}, {Oh}, {Parejko}, {Parley}, {Pascual},
  {Patil}, {Patil}, {Plunkett}, {Prochaska}, {Rastogi}, {Reddy Janga},
  {Sabater}, {Sakurikar}, {Seifert}, {Sherbert}, {Sherwood-Taylor}, {Shih},
  {Sick}, {Silbiger}, {Singanamalla}, {Singer}, {Sladen}, {Sooley},
  {Sornarajah}, {Streicher}, {Teuben}, {Thomas}, {Tremblay}, {Turner},
  {Terr{\'o}n}, {van Kerkwijk}, {de la Vega}, {Watkins}, {Weaver}, {Whitmore},
  {Woillez}, {Zabalza}, \& {Astropy Contributors}}]{astropy:2018}
{Astropy Collaboration}, {Price-Whelan}, A.~M., {Sip{\H{o}}cz}, B.~M., {et~al.}
  2018, \aj, 156, 123, \dodoi{10.3847/1538-3881/aabc4f}

\bibitem[{{Astropy Collaboration} {et~al.}(2022){Astropy Collaboration},
  {Price-Whelan}, {Lim}, {Earl}, {Starkman}, {Bradley}, {Shupe}, {Patil},
  {Corrales}, {Brasseur}, {N{"o}the}, {Donath}, {Tollerud}, {Morris},
  {Ginsburg}, {Vaher}, {Weaver}, {Tocknell}, {Jamieson}, {van Kerkwijk},
  {Robitaille}, {Merry}, {Bachetti}, {G{"u}nther}, {Aldcroft},
  {Alvarado-Montes}, {Archibald}, {B{'o}di}, {Bapat}, {Barentsen}, {Baz{'a}n},
  {Biswas}, {Boquien}, {Burke}, {Cara}, {Cara}, {Conroy}, {Conseil}, {Craig},
  {Cross}, {Cruz}, {D'Eugenio}, {Dencheva}, {Devillepoix}, {Dietrich},
  {Eigenbrot}, {Erben}, {Ferreira}, {Foreman-Mackey}, {Fox}, {Freij}, {Garg},
  {Geda}, {Glattly}, {Gondhalekar}, {Gordon}, {Grant}, {Greenfield}, {Groener},
  {Guest}, {Gurovich}, {Handberg}, {Hart}, {Hatfield-Dodds}, {Homeier},
  {Hosseinzadeh}, {Jenness}, {Jones}, {Joseph}, {Kalmbach}, {Karamehmetoglu},
  {Ka{l}uszy{'n}ski}, {Kelley}, {Kern}, {Kerzendorf}, {Koch}, {Kulumani},
  {Lee}, {Ly}, {Ma}, {MacBride}, {Maljaars}, {Muna}, {Murphy}, {Norman},
  {O'Steen}, {Oman}, {Pacifici}, {Pascual}, {Pascual-Granado}, {Patil},
  {Perren}, {Pickering}, {Rastogi}, {Roulston}, {Ryan}, {Rykoff}, {Sabater},
  {Sakurikar}, {Salgado}, {Sanghi}, {Saunders}, {Savchenko}, {Schwardt},
  {Seifert-Eckert}, {Shih}, {Jain}, {Shukla}, {Sick}, {Simpson},
  {Singanamalla}, {Singer}, {Singhal}, {Sinha}, {Sip{H{o}}cz}, {Spitler},
  {Stansby}, {Streicher}, {{{S}}umak}, {Swinbank}, {Taranu}, {Tewary},
  {Tremblay}, {Val-Borro}, {Van Kooten}, {Vasovi{'c}}, {Verma}, {de Miranda
  Cardoso}, {Williams}, {Wilson}, {Winkel}, {Wood-Vasey}, {Xue}, {Yoachim},
  {Zhang}, {Zonca}, \& {Astropy Project Contributors}}]{astropy:2022}
{Astropy Collaboration}, {Price-Whelan}, A.~M., {Lim}, P.~L., {et~al.} 2022,
  apj, 935, 167, \dodoi{10.3847/1538-4357/ac7c74}

\bibitem[{{Barcos-Mu{\~n}oz} {et~al.}(2018){Barcos-Mu{\~n}oz}, {Aalto},
  {Thompson}, {Sakamoto}, {Mart{\'\i}n}, {Leroy}, {Privon}, {Evans}, \&
  {Kepley}}]{barcos-munoz18}
{Barcos-Mu{\~n}oz}, L., {Aalto}, S., {Thompson}, T.~A., {et~al.} 2018, \apjl,
  853, L28, \dodoi{10.3847/2041-8213/aaa28d}

\bibitem[{{Bertola} {et~al.}(1991){Bertola}, {Bettoni}, {Danziger}, {Sadler},
  {Sparke}, \& {de Zeeuw}}]{bertola91}
{Bertola}, F., {Bettoni}, D., {Danziger}, J., {et~al.} 1991, \apj, 373, 369,
  \dodoi{10.1086/170058}

\bibitem[{{Bianchin} {et~al.}(2022){Bianchin}, {Riffel}, {Storchi-Bergmann},
  {Riffel}, {Ruschel-Dutra}, {Harrison}, {Dahmer-Hahn}, {Mainieri},
  {Sch{\"o}nell}, \& {Dametto}}]{bianchin22}
{Bianchin}, M., {Riffel}, R.~A., {Storchi-Bergmann}, T., {et~al.} 2022, \mnras,
  510, 639, \dodoi{10.1093/mnras/stab3468}

\bibitem[{{B{\^\i}rzan} {et~al.}(2008){B{\^\i}rzan}, {McNamara}, {Nulsen},
  {Carilli}, \& {Wise}}]{birzan08}
{B{\^\i}rzan}, L., {McNamara}, B.~R., {Nulsen}, P.~E.~J., {Carilli}, C.~L., \&
  {Wise}, M.~W. 2008, \apj, 686, 859, \dodoi{10.1086/591416}

\bibitem[{{Bohn} {et~al.}(2023){Bohn}, {Inami}, {Diaz-Santos}, {Armus},
  {Linden}, {U}, {Surace}, {Larson}, {Evans}, {Hoshioka}, {Lai}, {Song},
  {Mazzarella}, {Barcos-Munoz}, {Charmandaris}, {Howell}, {Medling}, {Privon},
  {Rich}, {Stierwalt}, {Aalto}, {B{\"o}ker}, {Brown}, {Iwasawa}, {Malkan}, {van
  der Werf}, {Appleton}, {Hayward}, {Kemper}, {Law}, {Marshall}, {Murphy}, \&
  {Sanders}}]{bohn23}
{Bohn}, T., {Inami}, H., {Diaz-Santos}, T., {et~al.} 2023, \apjl, 942, L36,
  \dodoi{10.3847/2041-8213/acab61}

\bibitem[{{B{\"o}ker} {et~al.}(2022){B{\"o}ker}, {Arribas}, {L{\"u}tzgendorf},
  {Alves de Oliveira}, {Beck}, {Birkmann}, {Bunker}, {Charlot}, {de Marchi},
  {Ferruit}, {Giardino}, {Jakobsen}, {Kumari}, {L{\'o}pez-Caniego}, {Maiolino},
  {Manjavacas}, {Marston}, {Moseley}, {Muzerolle}, {Ogle}, {Pirzkal},
  {Rauscher}, {Rawle}, {Rix}, {Sabbi}, {Sargent}, {Sirianni}, {te Plate},
  {Valenti}, {Willott}, \& {Zeidler}}]{boker22}
{B{\"o}ker}, T., {Arribas}, S., {L{\"u}tzgendorf}, N., {et~al.} 2022, \aap,
  661, A82, \dodoi{10.1051/0004-6361/202142589}

\bibitem[{{Bouchet} {et~al.}(2015){Bouchet}, {Garc{\'\i}a-Mar{\'\i}n},
  {Lagage}, {Amiaux}, {Augu{\'e}res}, {Bauwens}, {Blommaert}, {Chen}, {Detre},
  {Dicken}, {Dubreuil}, {Galdemard}, {Gastaud}, {Glasse}, {Gordon}, {Gougnaud},
  {Guillard}, {Justtanont}, {Krause}, {Leboeuf}, {Longval}, {Martin}, {Mazy},
  {Moreau}, {Olofsson}, {Ray}, {Rees}, {Renotte}, {Ressler}, {Ronayette},
  {Salasca}, {Scheithauer}, {Sykes}, {Thelen}, {Wells}, {Wright}, \&
  {Wright}}]{bouchet15}
{Bouchet}, P., {Garc{\'\i}a-Mar{\'\i}n}, M., {Lagage}, P.~O., {et~al.} 2015,
  \pasp, 127, 612, \dodoi{10.1086/682254}

\bibitem[{{Bushouse} {et~al.}(2022){Bushouse}, {Eisenhamer}, {Dencheva},
  {Davies}, {Greenfield}, {Morrison}, {Hodge}, {Simon}, {Grumm}, {Droettboom},
  {Slavich}, {Sosey}, {Pauly}, {Miller}, {Jedrzejewski}, {Hack}, {Davis},
  {Crawford}, {Law}, {Gordon}, {Regan}, {Cara}, {MacDonald}, {Bradley},
  {Shanahan}, {Jamieson}, {Teodoro}, \& {Williams}}]{bushouse2022}
{Bushouse}, H., {Eisenhamer}, J., {Dencheva}, N., {et~al.} 2022, {JWST
  Calibration Pipeline}, 1.8.2,  Zenodo, \dodoi{10.5281/zenodo.7229890}

\bibitem[{{Cavagnolo} {et~al.}(2010){Cavagnolo}, {McNamara}, {Nulsen},
  {Carilli}, {Jones}, \& {B{\^\i}rzan}}]{cavagnolo10}
{Cavagnolo}, K.~W., {McNamara}, B.~R., {Nulsen}, P.~E.~J., {et~al.} 2010, \apj,
  720, 1066, \dodoi{10.1088/0004-637X/720/2/1066}

\bibitem[{{Clavijo-Boh{\'o}rquez} {et~al.}(2023){Clavijo-Boh{\'o}rquez}, {de
  Gouveia Dal Pino}, \& {Melioli}}]{clavijo23}
{Clavijo-Boh{\'o}rquez}, W.~E., {de Gouveia Dal Pino}, E.~M., \& {Melioli}, C.
  2023, arXiv e-prints, arXiv:2306.11494, \dodoi{10.48550/arXiv.2306.11494}

\bibitem[{{Colina} {et~al.}(2015){Colina}, {Piqueras L{\'o}pez}, {Arribas},
  {Riffel}, {Riffel}, {Rodriguez-Ardila}, {Pastoriza}, {Storchi-Bergmann},
  {Alonso-Herrero}, \& {Sales}}]{colina15}
{Colina}, L., {Piqueras L{\'o}pez}, J., {Arribas}, S., {et~al.} 2015, \aap,
  578, A48, \dodoi{10.1051/0004-6361/201425567}

\bibitem[{{Davies} {et~al.}(2004){Davies}, {Tacconi}, \& {Genzel}}]{davies04}
{Davies}, R.~I., {Tacconi}, L.~J., \& {Genzel}, R. 2004, \apj, 602, 148,
  \dodoi{10.1086/380995}

\bibitem[{{de Vaucouleurs} {et~al.}(1991){de Vaucouleurs}, {de Vaucouleurs},
  {Corwin}, {Buta}, {Paturel}, \& {Fouque}}]{RC3}
{de Vaucouleurs}, G., {de Vaucouleurs}, A., {Corwin}, Herold~G., J., {et~al.}
  1991, {Third Reference Catalogue of Bright Galaxies}

\bibitem[{{D{\'\i}az-Santos} {et~al.}(2007){D{\'\i}az-Santos},
  {Alonso-Herrero}, {Colina}, {Ryder}, \& {Knapen}}]{diaz-santos07}
{D{\'\i}az-Santos}, T., {Alonso-Herrero}, A., {Colina}, L., {Ryder}, S.~D., \&
  {Knapen}, J.~H. 2007, \apj, 661, 149, \dodoi{10.1086/513089}

\bibitem[{{Falstad} {et~al.}(2018){Falstad}, {Aalto}, {Mangum}, {Costagliola},
  {Gallagher}, {Gonz{\'a}lez-Alfonso}, {Sakamoto}, {K{\"o}nig}, {Muller},
  {Evans}, \& {Privon}}]{falstad18}
{Falstad}, N., {Aalto}, S., {Mangum}, J.~G., {et~al.} 2018, \aap, 609, A75,
  \dodoi{10.1051/0004-6361/201732088}

\bibitem[{{Feruglio} {et~al.}(2010){Feruglio}, {Maiolino}, {Piconcelli},
  {Menci}, {Aussel}, {Lamastra}, \& {Fiore}}]{feruglio10}
{Feruglio}, C., {Maiolino}, R., {Piconcelli}, E., {et~al.} 2010, \aap, 518,
  L155, \dodoi{10.1051/0004-6361/201015164}

\bibitem[{{Fiore} {et~al.}(2017){Fiore}, {Feruglio}, {Shankar}, {Bischetti},
  {Bongiorno}, {Brusa}, {Carniani}, {Cicone}, {Duras}, {Lamastra}, {Mainieri},
  {Marconi}, {Menci}, {Maiolino}, {Piconcelli}, {Vietri}, \&
  {Zappacosta}}]{fiore17}
{Fiore}, F., {Feruglio}, C., {Shankar}, F., {et~al.} 2017, \aap, 601, A143,
  \dodoi{10.1051/0004-6361/201629478}

\bibitem[{{Forbes} \& {Ward}(1993)}]{forbes93}
{Forbes}, D.~A., \& {Ward}, M.~J. 1993, \apj, 416, 150, \dodoi{10.1086/173221}

\bibitem[{{Garc{\'\i}a-Bernete} {et~al.}(2022){Garc{\'\i}a-Bernete},
  {Rigopoulou}, {Alonso-Herrero}, {Donnan}, {Roche}, {Pereira-Santaella},
  {Labiano}, {Peralta de Arriba}, {Izumi}, {Ramos Almeida}, {Shimizu},
  {H{\"o}nig}, {Garc{\'\i}a-Burillo}, {Rosario}, {Ward}, {Bellocchi}, {Hicks},
  {Fuller}, \& {Packham}}]{garcia-bernete22}
{Garc{\'\i}a-Bernete}, I., {Rigopoulou}, D., {Alonso-Herrero}, A., {et~al.}
  2022, \aap, 666, L5, \dodoi{10.1051/0004-6361/202244806}

\bibitem[{{Gonz{\'a}lez-Alfonso} {et~al.}(2021){Gonz{\'a}lez-Alfonso},
  {Pereira-Santaella}, {Fischer}, {Garc{\'\i}a-Burillo}, {Yang},
  {Alonso-Herrero}, {Colina}, {Ashby}, {Smith}, {Rico-Villas},
  {Mart{\'\i}n-Pintado}, {Cazzoli}, \& {Stewart}}]{gonzalez-alfonso21}
{Gonz{\'a}lez-Alfonso}, E., {Pereira-Santaella}, M., {Fischer}, J., {et~al.}
  2021, \aap, 645, A49, \dodoi{10.1051/0004-6361/202039047}

\bibitem[{{Harrison} {et~al.}(2014){Harrison}, {Alexander}, {Mullaney}, \&
  {Swinbank}}]{harrison14}
{Harrison}, C.~M., {Alexander}, D.~M., {Mullaney}, J.~R., \& {Swinbank}, A.~M.
  2014, \mnras, 441, 3306, \dodoi{10.1093/mnras/stu515}

\bibitem[{{Hekatelyne} {et~al.}(2020){Hekatelyne}, {Riffel},
  {Storchi-Bergmann}, {Kharb}, {Robinson}, {Sales}, \&
  {Cassanta}}]{hekatelyne20}
{Hekatelyne}, C., {Riffel}, R.~A., {Storchi-Bergmann}, T., {et~al.} 2020,
  \mnras, 498, 2632, \dodoi{10.1093/mnras/staa2479}

\bibitem[{{Hekatelyne} {et~al.}(2018){Hekatelyne}, {Riffel}, {Sales},
  {Robinson}, {Storchi-Bergmann}, {Kharb}, {Gallimore}, {Baum}, \&
  {O'Dea}}]{hekatelyne18}
{Hekatelyne}, C., {Riffel}, R.~A., {Sales}, D., {et~al.} 2018, \mnras, 479,
  3966, \dodoi{10.1093/mnras/sty1606}

\bibitem[{{Hicks} \& {Malkan}(2008)}]{hicks08}
{Hicks}, E. K.~S., \& {Malkan}, M.~A. 2008, \apjs, 174, 31,
  \dodoi{10.1086/521650}

\bibitem[{{Hollenbach} \& {McKee}(1989)}]{hollenbach89}
{Hollenbach}, D., \& {McKee}, C.~F. 1989, \apj, 342, 306,
  \dodoi{10.1086/167595}

\bibitem[{Hunter(2007)}]{hunter07}
Hunter, J.~D. 2007, Computing in Science \& Engineering, 9, 90,
  \dodoi{10.1109/MCSE.2007.55}

\bibitem[{{Hutchison} {et~al.}(2023){Hutchison}, {Welch}, {Rigby}, {Olivier},
  {Birkin}, {Phadke}, {Khullar}, {Rauscher}, {Sharon}, {Aravena}, {Bayliss},
  {Elicker}, {Kim}, {Solimano}, {Vieira}, \& {Vizgan}}]{hutchison23}
{Hutchison}, T.~A., {Welch}, B.~D., {Rigby}, J.~R., {et~al.} 2023, arXiv
  e-prints, arXiv:2312.12518, \dodoi{10.48550/arXiv.2312.12518}

\bibitem[{{Jakobsen} {et~al.}(2022){Jakobsen}, {Ferruit}, {Alves de Oliveira},
  {Arribas}, {Bagnasco}, {Barho}, {Beck}, {Birkmann}, {B{\"o}ker}, {Bunker},
  {Charlot}, {de Jong}, {de Marchi}, {Ehrenwinkler}, {Falcolini}, {Fels},
  {Franx}, {Franz}, {Funke}, {Giardino}, {Gnata}, {Holota}, {Honnen}, {Jensen},
  {Jentsch}, {Johnson}, {Jollet}, {Karl}, {Kling}, {K{\"o}hler}, {Kolm},
  {Kumari}, {Lander}, {Lemke}, {L{\'o}pez-Caniego}, {L{\"u}tzgendorf},
  {Maiolino}, {Manjavacas}, {Marston}, {Maschmann}, {Maurer}, {Messerschmidt},
  {Moseley}, {Mosner}, {Mott}, {Muzerolle}, {Pirzkal}, {Pittet}, {Plitzke},
  {Posselt}, {Rapp}, {Rauscher}, {Rawle}, {Rix}, {R{\"o}del}, {Rumler},
  {Sabbi}, {Salvignol}, {Schmid}, {Sirianni}, {Smith}, {Strada}, {te Plate},
  {Valenti}, {Wettemann}, {Wiehe}, {Wiesmayer}, {Willott}, {Wright}, {Zeidler},
  \& {Zincke}}]{jakobsen22}
{Jakobsen}, P., {Ferruit}, P., {Alves de Oliveira}, C., {et~al.} 2022, \aap,
  661, A80, \dodoi{10.1051/0004-6361/202142663}

\bibitem[{{Kakkad} {et~al.}(2020){Kakkad}, {Mainieri}, {Vietri}, {Carniani},
  {Harrison}, {Perna}, {Scholtz}, {Circosta}, {Cresci}, {Husemann},
  {Bischetti}, {Feruglio}, {Fiore}, {Marconi}, {Padovani}, {Brusa}, {Cicone},
  {Comastri}, {Lanzuisi}, {Mannucci}, {Menci}, {Netzer}, {Piconcelli},
  {Puglisi}, {Salvato}, {Schramm}, {Silverman}, {Vignali}, {Zamorani}, \&
  {Zappacosta}}]{kakkad20}
{Kakkad}, D., {Mainieri}, V., {Vietri}, G., {et~al.} 2020, \aap, 642, A147,
  \dodoi{10.1051/0004-6361/202038551}

\bibitem[{{Lai} {et~al.}(2022){Lai}, {Armus}, {U}, {D{\'\i}az-Santos},
  {Larson}, {Evans}, {Malkan}, {Appleton}, {Rich}, {M{\"u}ller-S{\'a}nchez},
  {Inami}, {Bohn}, {McKinney}, {Finnerty}, {Law}, {Linden}, {Medling},
  {Privon}, {Song}, {Stierwalt}, {van der Werf}, {Barcos-Mu{\~n}oz}, {Smith},
  {Togi}, {Aalto}, {B{\"o}ker}, {Charmandaris}, {Howell}, {Iwasawa}, {Kemper},
  {Mazzarella}, {Murphy}, {Brown}, {Hayward}, {Marshall}, {Sanders}, \&
  {Surace}}]{lai22}
{Lai}, T. S.~Y., {Armus}, L., {U}, V., {et~al.} 2022, \apjl, 941, L36,
  \dodoi{10.3847/2041-8213/ac9ebf}

\bibitem[{{Lai} {et~al.}(2023){Lai}, {Armus}, {Bianchin}, {D{\'\i}az-Santos},
  {Linden}, {Privon}, {Inami}, {U}, {Bohn}, {Evans}, {Larson}, {Hensley},
  {Smith}, {Malkan}, {Song}, {Stierwalt}, {van der Werf}, {McKinney}, {Aalto},
  {Buiten}, {Rich}, {Charmandaris}, {Appleton}, {Barcos-Mu{\~n}oz},
  {B{\"o}ker}, {Finnerty}, {Kader}, {Law}, {Medling}, {Brown}, {Hayward},
  {Howell}, {Iwasawa}, {Kemper}, {Marshall}, {Mazzarella},
  {M{\"u}ller-S{\'a}nchez}, {Murphy}, {Sanders}, \& {Surace}}]{lai23}
{Lai}, T. S.~Y., {Armus}, L., {Bianchin}, M., {et~al.} 2023, \apjl, 957, L26,
  \dodoi{10.3847/2041-8213/ad0387}

\bibitem[{{Landt} {et~al.}(2008){Landt}, {Bentz}, {Ward}, {Elvis}, {Peterson},
  {Korista}, \& {Karovska}}]{landt08}
{Landt}, H., {Bentz}, M.~C., {Ward}, M.~J., {et~al.} 2008, \apjs, 174, 282,
  \dodoi{10.1086/522373}

\bibitem[{{Larkin} {et~al.}(1998){Larkin}, {Armus}, {Knop}, {Soifer}, \&
  {Matthews}}]{larkin98}
{Larkin}, J.~E., {Armus}, L., {Knop}, R.~A., {Soifer}, B.~T., \& {Matthews}, K.
  1998, \apjs, 114, 59, \dodoi{10.1086/313063}

\bibitem[{{Lonsdale} {et~al.}(2003){Lonsdale}, {Lonsdale}, {Smith}, \&
  {Diamond}}]{lonsdale03}
{Lonsdale}, C.~J., {Lonsdale}, C.~J., {Smith}, H.~E., \& {Diamond}, P.~J. 2003,
  \apj, 592, 804, \dodoi{10.1086/375778}

\bibitem[{{Lu} {et~al.}(2021){Lu}, {Wang}, {Zhang}, {Huang}, {Xu}, {Xin}, {Yu},
  {Ding}, {Wang}, \& {Feng}}]{lu21}
{Lu}, K.-X., {Wang}, J.-G., {Zhang}, Z.-X., {et~al.} 2021, \apj, 918, 50,
  \dodoi{10.3847/1538-4357/ac0c78}

\bibitem[{{Lutz} {et~al.}(2020){Lutz}, {Sturm}, {Janssen}, {Veilleux}, {Aalto},
  {Cicone}, {Contursi}, {Davies}, {Feruglio}, {Fischer}, {Fluetsch},
  {Garcia-Burillo}, {Genzel}, {Gonz{\'a}lez-Alfonso}, {Graci{\'a}-Carpio},
  {Herrera-Camus}, {Maiolino}, {Schruba}, {Shimizu}, {Sternberg}, {Tacconi}, \&
  {Wei{\ss}}}]{lutz20}
{Lutz}, D., {Sturm}, E., {Janssen}, A., {et~al.} 2020, \aap, 633, A134,
  \dodoi{10.1051/0004-6361/201936803}

\bibitem[{{Markwardt}(2009)}]{markwardt09}
{Markwardt}, C.~B. 2009, in Astronomical Society of the Pacific Conference
  Series, Vol. 411, Astronomical Data Analysis Software and Systems XVIII, ed.
  D.~A. {Bohlender}, D.~{Durand}, \& P.~{Dowler}, 251,
  \dodoi{10.48550/arXiv.0902.2850}

\bibitem[{{Medling} {et~al.}(2015){Medling}, {U}, {Rich}, {Kewley}, {Armus},
  {Dopita}, {Max}, {Sanders}, \& {Sutherland}}]{medling15}
{Medling}, A.~M., {U}, V., {Rich}, J.~A., {et~al.} 2015, \mnras, 448, 2301,
  \dodoi{10.1093/mnras/stv081}

\bibitem[{{Medling} {et~al.}(2019){Medling}, {Privon}, {Barcos-Mu{\~n}oz},
  {Treister}, {Cicone}, {Messias}, {Sanders}, {Scoville}, {U}, {Armus},
  {Bauer}, {Chang}, {Comerford}, {Evans}, {Max}, {M{\"u}ller-S{\'a}nchez},
  {Nagar}, \& {Sheth}}]{medling19}
{Medling}, A.~M., {Privon}, G.~C., {Barcos-Mu{\~n}oz}, L., {et~al.} 2019,
  \apjl, 885, L21, \dodoi{10.3847/2041-8213/ab4db7}

\bibitem[{{Morganti} {et~al.}(2015){Morganti}, {Oosterloo}, {Oonk},
  {Frieswijk}, \& {Tadhunter}}]{morganti15}
{Morganti}, R., {Oosterloo}, T., {Oonk}, J.~B.~R., {Frieswijk}, W., \&
  {Tadhunter}, C. 2015, \aap, 580, A1, \dodoi{10.1051/0004-6361/201525860}

\bibitem[{Morganti {et~al.}(2016)Morganti, Veilleux, Oosterloo, Teng, \&
  Rupke}]{morganti16}
Morganti, R., Veilleux, S., Oosterloo, T., Teng, S.~H., \& Rupke, D. 2016,
  Astronomy \& Astrophysics, 593, A30, \dodoi{10.1051/0004-6361/201628978}

\bibitem[{{Motter} {et~al.}(2021){Motter}, {Riffel}, {Ricci}, {Riffel},
  {Storchi-Bergmann}, {Pastoriza}, {Rodriguez-Ardila}, {Ruschel-Dutra},
  {Dahmer-Hahn}, {Dametto}, \& {Diniz}}]{motter21}
{Motter}, J.~C., {Riffel}, R., {Ricci}, T.~V., {et~al.} 2021, \mnras, 506,
  4354, \dodoi{10.1093/mnras/stab1977}

\bibitem[{{Mukherjee} {et~al.}(2018){Mukherjee}, {Wagner}, {Bicknell},
  {Morganti}, {Oosterloo}, {Nesvadba}, \& {Sutherland}}]{mukherjee18}
{Mukherjee}, D., {Wagner}, A.~Y., {Bicknell}, G.~V., {et~al.} 2018, \mnras,
  476, 80, \dodoi{10.1093/mnras/sty067}

\bibitem[{{M{\"u}ller-S{\'a}nchez} {et~al.}(2011){M{\"u}ller-S{\'a}nchez},
  {Prieto}, {Hicks}, {Vives-Arias}, {Davies}, {Malkan}, {Tacconi}, \&
  {Genzel}}]{muller-sanchez11}
{M{\"u}ller-S{\'a}nchez}, F., {Prieto}, M.~A., {Hicks}, E.~K.~S., {et~al.}
  2011, \apj, 739, 69, \dodoi{10.1088/0004-637X/739/2/69}

\bibitem[{{Orienti} \& {Prieto}(2010)}]{orienti10}
{Orienti}, M., \& {Prieto}, M.~A. 2010, \mnras, 401, 2599,
  \dodoi{10.1111/j.1365-2966.2009.15837.x}

\bibitem[{{Osterbrock}(1977)}]{osterbrock77}
{Osterbrock}, D.~E. 1977, \apj, 215, 733, \dodoi{10.1086/155407}

\bibitem[{{Osterbrock} \& {Ferland}(2006)}]{osterbrock}
{Osterbrock}, D.~E., \& {Ferland}, G.~J. 2006, {Astrophysics of gaseous nebulae
  and active galactic nuclei}

\bibitem[{{Ott}(2012)}]{ott12}
{Ott}, T. 2012, {QFitsView: FITS file viewer}, Astrophysics Source Code
  Library, record ascl:1210.019.
\newblock \doeprint{1210.019}

\bibitem[{{Pereira-Santaella} {et~al.}(2020){Pereira-Santaella}, {Colina},
  {Garc{\'\i}a-Burillo}, {Gonz{\'a}lez-Alfonso}, {Alonso-Herrero}, {Arribas},
  {Cazzoli}, {Piqueras-L{\'o}pez}, {Rigopoulou}, \&
  {Usero}}]{pereira-santaella20}
{Pereira-Santaella}, M., {Colina}, L., {Garc{\'\i}a-Burillo}, S., {et~al.}
  2020, \aap, 643, A89, \dodoi{10.1051/0004-6361/202038838}

\bibitem[{{P{\'e}rez-Torres} {et~al.}(2021){P{\'e}rez-Torres}, {Mattila},
  {Alonso-Herrero}, {Aalto}, \& {Efstathiou}}]{perez-torres21}
{P{\'e}rez-Torres}, M., {Mattila}, S., {Alonso-Herrero}, A., {Aalto}, S., \&
  {Efstathiou}, A. 2021, \aapr, 29, 2, \dodoi{10.1007/s00159-020-00128-x}

\bibitem[{{Peterson} {et~al.}(2014){Peterson}, {Grier}, {Horne}, {Pogge},
  {Bentz}, {De Rosa}, {Denney}, {Martini}, {Sergeev}, {Kaspi}, {Minezaki},
  {Zu}, {Kochanek}, {Siverd}, {Shappee}, {Araya Salvo}, {Beatty}, {Bird},
  {Bord}, {Borman}, {Che}, {Chen}, {Cohen}, {Dietrich}, {Doroshenko}, {Drake},
  {Efimov}, {Free}, {Ginsburg}, {Henderson}, {King}, {Koshida}, {Mogren},
  {Molina}, {Mosquera}, {Motohara}, {Nazarov}, {Okhmat}, {Pejcha}, {Rafter},
  {Shields}, {Skowron}, {Skowron}, {Valluri}, {van Saders}, \&
  {Yoshii}}]{peterson14}
{Peterson}, B.~M., {Grier}, C.~J., {Horne}, K., {et~al.} 2014, \apj, 795, 149,
  \dodoi{10.1088/0004-637X/795/2/149}

\bibitem[{{Raimundo} {et~al.}(2019){Raimundo}, {Vestergaard}, {Koay},
  {Lawther}, {Casasola}, \& {Peterson}}]{raimundo19}
{Raimundo}, S.~I., {Vestergaard}, M., {Koay}, J.~Y., {et~al.} 2019, \mnras,
  486, 123, \dodoi{10.1093/mnras/stz852}

\bibitem[{{Reunanen} {et~al.}(2002){Reunanen}, {Kotilainen}, \&
  {Prieto}}]{reunanen02}
{Reunanen}, J., {Kotilainen}, J.~K., \& {Prieto}, M.~A. 2002, \mnras, 331, 154,
  \dodoi{10.1046/j.1365-8711.2002.05181.x}

\bibitem[{{Rich} {et~al.}(2014){Rich}, {Kewley}, \& {Dopita}}]{rich14}
{Rich}, J.~A., {Kewley}, L.~J., \& {Dopita}, M.~A. 2014, \apjl, 781, L12,
  \dodoi{10.1088/2041-8205/781/1/L12}

\bibitem[{{Rich} {et~al.}(2015){Rich}, {Kewley}, \& {Dopita}}]{rich15}
---. 2015, \apjs, 221, 28, \dodoi{10.1088/0067-0049/221/2/28}

\bibitem[{{Rich} {et~al.}(2012){Rich}, {Torrey}, {Kewley}, {Dopita}, \&
  {Rupke}}]{rich12}
{Rich}, J.~A., {Torrey}, P., {Kewley}, L.~J., {Dopita}, M.~A., \& {Rupke},
  D.~S.~N. 2012, \apj, 753, 5, \dodoi{10.1088/0004-637X/753/1/5}

\bibitem[{{Rieke} {et~al.}(2023){Rieke}, {Kelly}, {Misselt}, {Stansberry},
  {Boyer}, {Beatty}, {Egami}, {Florian}, {Greene}, {Hainline}, {Leisenring},
  {Roellig}, {Schlawin}, {Sun}, {Tinnin}, {Williams}, {Willmer}, {Wilson},
  {Clark}, {Rohrbach}, {Brooks}, {Canipe}, {Correnti}, {DiFelice}, {Gennaro},
  {Girard}, {Hartig}, {Hilbert}, {Koekemoer}, {Nikolov}, {Pirzkal}, {Rest},
  {Robberto}, {Sunnquist}, {Telfer}, {Wu}, {Ferry}, {Lewis}, {Baum},
  {Beichman}, {Doyon}, {Dressler}, {Eisenstein}, {Ferrarese}, {Hodapp},
  {Horner}, {Jaffe}, {Johnstone}, {Krist}, {Martin}, {McCarthy}, {Meyer},
  {Rieke}, {Trauger}, \& {Young}}]{rieke23}
{Rieke}, M.~J., {Kelly}, D.~M., {Misselt}, K., {et~al.} 2023, \pasp, 135,
  028001, \dodoi{10.1088/1538-3873/acac53}

\bibitem[{{Riffel} {et~al.}(2013){Riffel}, {Rodr{\'\i}guez-Ardila}, {Aleman},
  {Brotherton}, {Pastoriza}, {Bonatto}, \& {Dors}}]{riffel13}
{Riffel}, R., {Rodr{\'\i}guez-Ardila}, A., {Aleman}, I., {et~al.} 2013, \mnras,
  430, 2002, \dodoi{10.1093/mnras/stt026}

\bibitem[{{Riffel} {et~al.}(2021{\natexlab{a}}){Riffel}, {Bianchin}, {Riffel},
  {Storchi-Bergmann}, {Sch{\"o}nell}, {Dahmer-Hahn}, {Dametto}, \&
  {Diniz}}]{riffel21}
{Riffel}, R.~A., {Bianchin}, M., {Riffel}, R., {et~al.} 2021{\natexlab{a}},
  \mnras, 503, 5161, \dodoi{10.1093/mnras/stab788}

\bibitem[{{Riffel} {et~al.}(2023{\natexlab{a}}){Riffel}, {Riffel}, {Bianchin},
  {Storchi-Bergmann}, {Souza-Oliveira}, \& {Zakamska}}]{riffelugc}
{Riffel}, R.~A., {Riffel}, R., {Bianchin}, M., {et~al.} 2023{\natexlab{a}},
  \mnras, 521, 3260, \dodoi{10.1093/mnras/stad776}

\bibitem[{{Riffel} {et~al.}(2015){Riffel}, {Storchi-Bergmann}, \&
  {Riffel}}]{riffel15}
{Riffel}, R.~A., {Storchi-Bergmann}, T., \& {Riffel}, R. 2015, \mnras, 451,
  3587, \dodoi{10.1093/mnras/stv1129}

\bibitem[{{Riffel} {et~al.}(2020){Riffel}, {Storchi-Bergmann}, {Zakamska}, \&
  {Riffel}}]{riffel20}
{Riffel}, R.~A., {Storchi-Bergmann}, T., {Zakamska}, N.~L., \& {Riffel}, R.
  2020, \mnras, 496, 4857, \dodoi{10.1093/mnras/staa1922}

\bibitem[{{Riffel} {et~al.}(2021{\natexlab{b}}){Riffel}, {Storchi-Bergmann},
  {Riffel}, {Bianchin}, {Zakamska}, {Ruschel-Dutra}, {Sch{\"o}nell}, {Rosario},
  {Rodriguez-Ardila}, {Fischer}, {Davies}, {Dametto}, {Dahmer-Hahn},
  {Crenshaw}, {Burtscher}, \& {Bentz}}]{riffel21_llp_exc}
{Riffel}, R.~A., {Storchi-Bergmann}, T., {Riffel}, R., {et~al.}
  2021{\natexlab{b}}, \mnras, 504, 3265, \dodoi{10.1093/mnras/stab998}

\bibitem[{{Riffel} {et~al.}(2023{\natexlab{b}}){Riffel}, {Storchi-Bergmann},
  {Riffel}, {Bianchin}, {Zakamska}, {Ruschel-Dutra}, {Bentz}, {Burtscher},
  {Crenshaw}, {Dahmer-Hahn}, {Dametto}, {Davies}, {Diniz}, {Fischer},
  {Harrison}, {Mainieri}, {Revalski}, {Rodriguez-Ardila}, {Rosario}, \&
  {Sch{\"o}nell}}]{riffelagnifskin}
---. 2023{\natexlab{b}}, \mnras, 521, 1832, \dodoi{10.1093/mnras/stad599}

\bibitem[{{Rigby} {et~al.}(2023){Rigby}, {Perrin}, {McElwain}, {Kimble},
  {Friedman}, {Lallo}, {Doyon}, {Feinberg}, {Ferruit}, {Glasse}, {Rieke},
  {Rieke}, {Wright}, {Willott}, {Colon}, {Milam}, {Neff}, {Stark}, {Valenti},
  {Abell}, {Abney}, {Abul-Huda}, {Acton}, {Adams}, {Adler}, {Aguilar}, {Ahmed},
  {Albert}, {Alberts}, {Aldridge}, {Allen}, {Altenburg},
  {{\'A}lvarez-M{\'a}rquez}, {Alves de Oliveira}, {Andersen}, {Anderson},
  {Anderson}, {Argyriou}, {Armstrong}, {Arribas}, {Artigau}, {Arvai},
  {Atkinson}, {Bacon}, {Bair}, {Banks}, {Barrientes}, {Barringer}, {Bartosik},
  {Bast}, {Baudoz}, {Beatty}, {Bechtold}, {Beck}, {Bergeron}, {Bergkoetter},
  {Bhatawdekar}, {Birkmann}, {Blazek}, {Blome}, {Boccaletti}, {B{\"o}ker},
  {Boia}, {Bonaventura}, {Bond}, {Bosley}, {Boucarut}, {Bourque}, {Bouwman},
  {Bower}, {Bowers}, {Boyer}, {Bradley}, {Brady}, {Braun}, {Breda},
  {Bresnahan}, {Bright}, {Britt}, {Bromenschenkel}, {Brooks}, {Brooks},
  {Brown}, {Brown}, {Brown}, {Bunker}, {Burger}, {Bushouse}, {Cale}, {Cameron},
  {Cameron}, {Canipe}, {Caplinger}, {Caputo}, {Cara}, {Carey}, {Carniani},
  {Carrasquilla}, {Carruthers}, {Case}, {Catherine}, {Chance}, {Chapman},
  {Charlot}, {Charlow}, {Chayer}, {Chen}, {Cherinka}, {Chichester}, {Chilton},
  {Chonis}, {Clampin}, {Clark}, {Clark}, {Coe}, {Coleman}, {Comber}, {Comeau},
  {Connolly}, {Cooper}, {Cooper}, {Coppock}, {Correnti}, {Cossou}, {Coulais},
  {Coyle}, {Cracraft}, {Curti}, {Cuturic}, {Davis}, {Davis}, {Dean}, {DeLisa},
  {deMeester}, {Dencheva}, {Dencheva}, {DePasquale}, {Deschenes}, {Hunor
  Detre}, {Diaz}, {Dicken}, {DiFelice}, {Dillman}, {Dixon}, {Doggett},
  {Donaldson}, {Douglas}, {DuPrie}, {Dupuis}, {Durning}, {Easmin}, {Eck},
  {Edeani}, {Egami}, {Ehrenwinkler}, {Eisenhamer}, {Eisenhower}, {Elie},
  {Elliott}, {Elliott}, {Ellis}, {Engesser}, {Espinoza}, {Etienne}, {Etxaluze},
  {Falini}, {Feeney}, {Ferry}, {Filippazzo}, {Fincham}, {Fix}, {Flagey},
  {Florian}, {Flynn}, {Fontanella}, {Ford}, {Forshay}, {Fox}, {Franz}, {Fu},
  {Fullerton}, {Galkin}, {Galyer}, {Garc{\'\i}a Mar{\'\i}n}, {Gardner},
  {Gardner}, {Garland}, {Garrett}, {Gasman}, {Gaspar}, {Gaudreau}, {Gauthier},
  {Geers}, {Geithner}, {Gennaro}, {Giardino}, {Girard}, {Giuliano},
  {Glassmire}, {Glauser}, {Glazer}, {Godfrey}, {Golimowski}, {Gollnitz},
  {Gong}, {Gonzaga}, {Gordon}, {Gordon}, {Goudfrooij}, {Greene}, {Greenhouse},
  {Grimaldi}, {Groebner}, {Grundy}, {Guillard}, {Gutman}, {Ha}, {Haderlein},
  {Hagedorn}, {Hainline}, {Haley}, {Hami}, {Hamilton}, {Hammel}, {Hansen},
  {Harkins}, {Harr}, {Hart}, {Hart}, {Hartig}, {Hashimoto}, {Haskins},
  {Hathaway}, {Havey}, {Hayden}, {Hecht}, {Heller-Boyer}, {Henriques}, {Henry},
  {Hermann}, {Hernandez}, {Hesman}, {Hicks}, {Hilbert}, {Hines}, {Hoffman},
  {Holfeltz}, {Holler}, {Hoppa}, {Hott}, {Howard}, {Howard}, {Hunter},
  {Hunter}, {Hurst}, {Husemann}, {Hustak}, {Ilinca Ignat}, {Illingworth},
  {Irish}, {Jackson}, {Jahromi}, {Jakobsen}, {James}, {James}, {Januszewski},
  {Jenkins}, {Jirdeh}, {Johnson}, {Johnson}, {Jones}, {Jones}, {Jones},
  {Jones}, {Jordan}, {Jordan}, {Jurczyk}, {Jurling}, {Kaleida}, {Kalmanson},
  {Kammerer}, {Kang}, {Kao}, {Karakla}, {Kavanagh}, {Kelly}, {Kendrew},
  {Kennedy}, {Kenny}, {Keski-kuha}, {Keyes}, {Kidwell}, {Kinzel}, {Kirk},
  {Kirkpatrick}, {Kirshenblat}, {Klaassen}, {Knapp}, {Knight}, {Knollenberg},
  {Koehler}, {Koekemoer}, {Kovacs}, {Kulp}, {Kumari}, {Kyprianou}, {La Massa},
  {Labador}, {Labiano}, {Lagage}, {Lajoie}, {Lallo}, {Lam}, {Lamb}, {Lambros},
  {Lampenfield}, {Langston}, {Larson}, {Law}, {Lawrence}, {Lee}, {Leisenring},
  {Lepo}, {Leveille}, {Levenson}, {Levine}, {Levy}, {Lewis}, {Lewis},
  {Libralato}, {Lightsey}, {Link}, {Liu}, {Lo}, {Lockwood}, {Logue}, {Long},
  {Long}, {Loomis}, {Lopez-Caniego}, {Lorenzo Alvarez}, {Love-Pruitt}, {Lucy},
  {Luetzgendorf}, {Maghami}, {Maiolino}, {Major}, {Malla}, {Malumuth},
  {Manjavacas}, {Mannfolk}, {Marrione}, {Marston}, {Martel}, {Maschmann},
  {Masci}, {Masciarelli}, {Maszkiewicz}, {Mather}, {McKenzie}, {McLean},
  {McMaster}, {Melbourne}, {Mel{\'e}ndez}, {Menzel}, {Merz}, {Meyett}, {Meza},
  {Miskey}, {Misselt}, {Moller}, {Morrison}, {Morse}, {Moseley}, {Mosier},
  {Mountain}, {Mueckay}, {Mueller}, {Mullally}, {Murphy}, {Murray}, {Murray},
  {Mustelier}, {Muzerolle}, {Mycroft}, {Myers}, {Myrick}, {Nanavati}, {Nance},
  {Nayak}, {Naylor}, {Nelan}, {Nickson}, {Nielson}, {Nieto-Santisteban},
  {Nikolov}, {Noriega-Crespo}, {O'Shaughnessy}, {O'Sullivan}, {Ochs}, {Ogle},
  {Oleszczuk}, {Olmsted}, {Osborne}, {Ottens}, {Owens}, {Pacifici}, {Pagan},
  {Page}, {Park}, {Parrish}, {Patapis}, {Paul}, {Pauly}, {Pavlovsky}, {Pedder},
  {Peek}, {Pena-Guerrero}, {Penanen}, {Perez}, {Perna}, {Perriello},
  {Phillips}, {Pietraszkiewicz}, {Pinaud}, {Pirzkal}, {Pitman}, {Piwowar},
  {Platais}, {Player}, {Plesha}, {Pollizi}, {Polster}, {Pontoppidan},
  {Porterfield}, {Proffitt}, {Pueyo}, {Pulliam}, {Quirt}, {Quispe Neira},
  {Ramos Alarcon}, {Ramsay}, {Rapp}, {Rapp}, {Rauscher}, {Ravindranath},
  {Rawle}, {Regan}, {Reichard}, {Reis}, {Ressler}, {Rest}, {Reynolds}, {Rhue},
  {Richon}, {Rickman}, {Ridgaway}, {Ritchie}, {Rix}, {Robberto}, {Robinson},
  {Robinson}, {Robinson}, {Rock}, {Rodriguez}, {Rodriguez Del Pino}, {Roellig},
  {Rohrbach}, {Roman}, {Romelfanger}, {Rose}, {Roteliuk}, {Roth}, {Rothwell},
  {Rowlands}, {Roy}, {Royer}, {Royle}, {Rui}, {Rumler}, {Runnels}, {Russ},
  {Rustamkulov}, {Ryden}, {Ryer}, {Sabata}, {Sabatke}, {Sabbi}, {Samuelson},
  {Sapp}, {Sappington}, {Sargent}, {Sauer}, {Scheithauer}, {Schlawin},
  {Schlitz}, {Schmitz}, {Schneider}, {Schreiber}, {Schulze}, {Schwab}, {Scott},
  {Sembach}, {Shanahan}, {Shaughnessy}, {Shaw}, {Shawger}, {Shay}, {Sheehan},
  {Shen}, {Sherman}, {Shiao}, {Shih}, {Shivaei}, {Sienkiewicz}, {Sing},
  {Sirianni}, {Sivaramakrishnan}, {Skipper}, {Sloan}, {Slocum}, {Slowinski},
  {Smith}, {Smith}, {Smith}, {Smith}, {Snyder}, {Soh}, {Sohn}, {Soto},
  {Spencer}, {Stallcup}, {Stansberry}, {Starr}, {Starr}, {Stewart},
  {Stiavelli}, {Straughn}, {Strickland}, {Stys}, {Summers}, {Sun}, {Sunnquist},
  {Swade}, {Swam}, {Swaters}, {Swoish}, {Taylor}, {Taylor}, {Te Plate}, {Tea},
  {Teague}, {Telfer}, {Temim}, {Thatte}, {Thompson}, {Thompson}, {Thomson},
  {Tikkanen}, {Tippet}, {Todd}, {Toolan}, {Tran}, {Trejo}, {Truong},
  {Tsukamoto}, {Tustain}, {Tyra}, {Ubeda}, {Underwood}, {Uzzo}, {Van Campen},
  {Vandal}, {Vandenbussche}, {Vila}, {Volk}, {Wahlgren}, {Waldman}, {Walker},
  {Wander}, {Warfield}, {Warner}, {Wasiak}, {Watkins}, {Weaver}, {Weilert},
  {Weiser}, {Weiss}, {Weissman}, {Welty}, {West}, {Wheate}, {Wheatley},
  {Wheeler}, {White}, {Whiteaker}, {Whitehouse}, {Whiteleather}, {Whitman},
  {Williams}, {Willmer}, {Willoughby}, {Wilson}, {Wirth}, {Wislowski}, {Wolf},
  {Wolfe}, {Wolff}, {Workman}, {Wright}, {Wu}, {Wu}, {Wymer}, {Yates},
  {Yeager}, {Yeates}, {Yerger}, {Yoon}, {Young}, {Yu}, {Zak}, {Zeidler},
  {Zhou}, {Zielinski}, {Zincke}, \& {Zonak}}]{rigby23}
{Rigby}, J., {Perrin}, M., {McElwain}, M., {et~al.} 2023, \pasp, 135, 048001,
  \dodoi{10.1088/1538-3873/acb293}

\bibitem[{{Robleto-Or{\'u}s} {et~al.}(2021){Robleto-Or{\'u}s},
  {Torres-Papaqui}, {Longinotti}, {Ortega-Minakata}, {S{\'a}nchez},
  {Ascasibar}, {Bellocchi}, {Galbany}, {Chow-Mart{\'\i}nez}, {Trejo-Alonso},
  {Morales-Vargas}, \& {Romero-Cruz}}]{robleto-orus21}
{Robleto-Or{\'u}s}, A.~C., {Torres-Papaqui}, J.~P., {Longinotti}, A.~L.,
  {et~al.} 2021, \apjl, 906, L6, \dodoi{10.3847/2041-8213/abd32f}

\bibitem[{{Rodr{\'\i}guez-Ardila} {et~al.}(2004){Rodr{\'\i}guez-Ardila},
  {Pastoriza}, {Viegas}, {Sigut}, \& {Pradhan}}]{alberto04}
{Rodr{\'\i}guez-Ardila}, A., {Pastoriza}, M.~G., {Viegas}, S., {Sigut},
  T.~A.~A., \& {Pradhan}, A.~K. 2004, \aap, 425, 457,
  \dodoi{10.1051/0004-6361:20034285}

\bibitem[{{Rodr{\'\i}guez-Ardila} {et~al.}(2006){Rodr{\'\i}guez-Ardila},
  {Prieto}, {Viegas}, \& {Gruenwald}}]{rodriguez-ardila06}
{Rodr{\'\i}guez-Ardila}, A., {Prieto}, M.~A., {Viegas}, S., \& {Gruenwald}, R.
  2006, \apj, 653, 1098, \dodoi{10.1086/508864}

\bibitem[{{Rodr{\'\i}guez-Ardila} {et~al.}(2005){Rodr{\'\i}guez-Ardila},
  {Riffel}, \& {Pastoriza}}]{alberto05}
{Rodr{\'\i}guez-Ardila}, A., {Riffel}, R., \& {Pastoriza}, M.~G. 2005, \mnras,
  364, 1041, \dodoi{10.1111/j.1365-2966.2005.09638.x}

\bibitem[{{Rosenberg} {et~al.}(2012){Rosenberg}, {van der Werf}, \&
  {Israel}}]{rosenberg12}
{Rosenberg}, M.~J.~F., {van der Werf}, P.~P., \& {Israel}, F.~P. 2012, \aap,
  540, A116, \dodoi{10.1051/0004-6361/201218772}

\bibitem[{{Rupke} \& {Veilleux}(2011)}]{rupke11}
{Rupke}, D. S.~N., \& {Veilleux}, S. 2011, \apjl, 729, L27,
  \dodoi{10.1088/2041-8205/729/2/L27}

\bibitem[{{Ruschel-Dutra} \& {Dall'Agnol De Oliveira}(2020)}]{ruschel-dutra20}
{Ruschel-Dutra}, D., \& {Dall'Agnol De Oliveira}, B. 2020, {danielrd6/ifscube:
  Modeling}, v1.1, Zenodo,  Zenodo, \dodoi{10.5281/zenodo.4065550}

\bibitem[{{Ruschel-Dutra} {et~al.}(2021){Ruschel-Dutra}, {Storchi-Bergmann},
  {Schnorr-M{\"u}ller}, {Riffel}, {Dall'Agnol de Oliveira}, {Lena}, {Robinson},
  {Nagar}, \& {Elvis}}]{ruschel-dutra21}
{Ruschel-Dutra}, D., {Storchi-Bergmann}, T., {Schnorr-M{\"u}ller}, A., {et~al.}
  2021, \mnras, 507, 74, \dodoi{10.1093/mnras/stab2058}

\bibitem[{{Sanders} \& {Mirabel}(1996)}]{sanders96}
{Sanders}, D.~B., \& {Mirabel}, I.~F. 1996, \araa, 34, 749,
  \dodoi{10.1146/annurev.astro.34.1.749}

\bibitem[{{Schnorr-M{\"u}ller} {et~al.}(2017){Schnorr-M{\"u}ller},
  {Storchi-Bergmann}, {Ferrari}, \& {Nagar}}]{schnorr-muller17}
{Schnorr-M{\"u}ller}, A., {Storchi-Bergmann}, T., {Ferrari}, F., \& {Nagar},
  N.~M. 2017, \mnras, 466, 4370, \dodoi{10.1093/mnras/stx018}

\bibitem[{{Schnorr-M{\"u}ller} {et~al.}(2016){Schnorr-M{\"u}ller},
  {Storchi-Bergmann}, {Robinson}, {Lena}, \& {Nagar}}]{schnorr-muller16}
{Schnorr-M{\"u}ller}, A., {Storchi-Bergmann}, T., {Robinson}, A., {Lena}, D.,
  \& {Nagar}, N.~M. 2016, \mnras, 457, 972, \dodoi{10.1093/mnras/stw037}

\bibitem[{{Simpson} {et~al.}(1996){Simpson}, {Forbes}, {Baker}, \&
  {Ward}}]{simpson96}
{Simpson}, C., {Forbes}, D.~A., {Baker}, A.~C., \& {Ward}, M.~J. 1996, \mnras,
  283, 777, \dodoi{10.1093/mnras/283.3.777}

\bibitem[{{Song} {et~al.}(2021){Song}, {Linden}, {Evans}, {Barcos-Mu{\~n}oz},
  {Privon}, {Yoon}, {Murphy}, {Larson}, {D{\'\i}az-Santos}, {Armus},
  {Mazzarella}, {Howell}, {Inami}, {Torres-Alb{\`a}}, {U}, {Charmandaris},
  {McKinney}, {Kunneriath}, \& {Momjian}}]{song21}
{Song}, Y., {Linden}, S.~T., {Evans}, A.~S., {et~al.} 2021, \apj, 916, 73,
  \dodoi{10.3847/1538-4357/ac05c2}

\bibitem[{{Song} {et~al.}(2022){Song}, {Linden}, {Evans}, {Barcos-Mu{\~n}oz},
  {Murphy}, {Momjian}, {D{\'\i}az-Santos}, {Larson}, {Privon}, {Huang},
  {Armus}, {Mazzarella}, {U}, {Inami}, {Charmandaris}, {Ricci}, {Emig},
  {McKinney}, {Yoon}, {Kunneriath}, {Lai}, {Rodas-Quito}, {Saravia}, {Gao},
  {Meynardie}, \& {Sanders}}]{song22}
---. 2022, \apj, 940, 52, \dodoi{10.3847/1538-4357/ac923b}

\bibitem[{{Storchi-Bergmann} {et~al.}(2007){Storchi-Bergmann}, {Dors},
  {Riffel}, {Fathi}, {Axon}, {Robinson}, {Marconi}, \& {{\"O}stlin}}]{thaisa07}
{Storchi-Bergmann}, T., {Dors}, Oli~L., J., {Riffel}, R.~A., {et~al.} 2007,
  \apj, 670, 959, \dodoi{10.1086/521918}

\bibitem[{{Storchi-Bergmann} {et~al.}(2009){Storchi-Bergmann}, {McGregor},
  {Riffel}, {Sim{\~o}es Lopes}, {Beck}, \& {Dopita}}]{storchi-bergmann09}
{Storchi-Bergmann}, T., {McGregor}, P.~J., {Riffel}, R.~A., {et~al.} 2009,
  \mnras, 394, 1148, \dodoi{10.1111/j.1365-2966.2009.14388.x}

\bibitem[{{Su} {et~al.}(2023){Su}, {Mahony}, {Gu}, {Sadler}, {Curran},
  {Allison}, {Yoon}, {Aditya}, {Chandola}, {Chen}, {Moss}, {Wu}, {Shao}, {Liu},
  {Glowacki}, {Whiting}, \& {Weng}}]{su23}
{Su}, R., {Mahony}, E.~K., {Gu}, M., {et~al.} 2023, \mnras, 520, 5712,
  \dodoi{10.1093/mnras/stad370}

\bibitem[{{U} {et~al.}(2013){U}, {Medling}, {Sanders}, {Max}, {Armus},
  {Iwasawa}, {Evans}, {Kewley}, \& {Fazio}}]{u13}
{U}, V., {Medling}, A., {Sanders}, D., {et~al.} 2013, \apj, 775, 115,
  \dodoi{10.1088/0004-637X/775/2/115}

\bibitem[{{U} {et~al.}(2019){U}, {Medling}, {Inami}, {Armus},
  {D{\'\i}az-Santos}, {Charmandaris}, {Howell}, {Stierwalt}, {Privon},
  {Linden}, {Sanders}, {Max}, {Evans}, {Barcos-Mu{\~n}oz}, {Chiang},
  {Appleton}, {Canalizo}, {Fazio}, {Iwasawa}, {Larson}, {Mazzarella}, {Murphy},
  {Rich}, \& {Surace}}]{u19}
{U}, V., {Medling}, A.~M., {Inami}, H., {et~al.} 2019, \apj, 871, 166,
  \dodoi{10.3847/1538-4357/aaf1c2}

\bibitem[{{U} {et~al.}(2022){U}, {Lai}, {Bianchin}, {Remigio}, {Armus},
  {Larson}, {D{\'\i}az-Santos}, {Evans}, {Stierwalt}, {Law}, {Malkan},
  {Linden}, {Song}, {van der Werf}, {Gao}, {Privon}, {Medling},
  {Barcos-Mu{\~n}oz}, {Hayward}, {Inami}, {Rich}, {Aalto}, {Appleton}, {Bohn},
  {B{\"o}ker}, {Brown}, {Charmandaris}, {Finnerty}, {Howell}, {Iwasawa},
  {Kemper}, {Marshall}, {Mazzarella}, {McKinney}, {Muller-Sanchez}, {Murphy},
  {Sanders}, \& {Surace}}]{u22c}
{U}, V., {Lai}, T., {Bianchin}, M., {et~al.} 2022, \apjl, 940, L5,
  \dodoi{10.3847/2041-8213/ac961c}

\bibitem[{{Veilleux} {et~al.}(2017){Veilleux}, {Bolatto}, {Tombesi},
  {Mel{\'e}ndez}, {Sturm}, {Gonz{\'a}lez-Alfonso}, {Fischer}, \&
  {Rupke}}]{veilleux17}
{Veilleux}, S., {Bolatto}, A., {Tombesi}, F., {et~al.} 2017, \apj, 843, 18,
  \dodoi{10.3847/1538-4357/aa767d}

\bibitem[{{Venturi} {et~al.}(2021){Venturi}, {Cresci}, {Marconi}, {Mingozzi},
  {Nardini}, {Carniani}, {Mannucci}, {Marasco}, {Maiolino}, {Perna},
  {Treister}, {Bland-Hawthorn}, \& {Gallimore}}]{venturi21}
{Venturi}, G., {Cresci}, G., {Marconi}, A., {et~al.} 2021, \aap, 648, A17,
  \dodoi{10.1051/0004-6361/202039869}

\bibitem[{Virtanen {et~al.}(2020)Virtanen, Gommers, Oliphant, Haberland, Reddy,
  Cournapeau, Burovski, Peterson, Weckesser, Bright, {van der Walt}, Brett,
  Wilson, Millman, Mayorov, Nelson, Jones, Kern, Larson, Carey, Polat, Feng,
  Moore, {VanderPlas}, Laxalde, Perktold, Cimrman, Henriksen, Quintero, Harris,
  Archibald, Ribeiro, Pedregosa, {van Mulbregt}, \& {SciPy 1.0
  Contributors}}]{SciPy}
Virtanen, P., Gommers, R., Oliphant, T.~E., {et~al.} 2020, Nature Methods, 17,
  261, \dodoi{10.1038/s41592-019-0686-2}

\bibitem[{{Xu} {et~al.}(2014){Xu}, {Cao}, {Lu}, {Gao}, {van der Werf}, {Evans},
  {Mazzarella}, {Chu}, {Haan}, {Diaz-Santos}, {Meijerink}, {Zhao}, {Appleton},
  {Armus}, {Charmandaris}, {Lord}, {Murphy}, {Sanders}, {Schulz}, \&
  {Stierwalt}}]{xu14}
{Xu}, C.~K., {Cao}, C., {Lu}, N., {et~al.} 2014, \apj, 787, 48,
  \dodoi{10.1088/0004-637X/787/1/48}

\bibitem[{{Xu} \& {Wang}(2022)}]{xu22}
{Xu}, X., \& {Wang}, J. 2022, \apj, 933, 110, \dodoi{10.3847/1538-4357/ac7222}

\bibitem[{{Zhang} \& {Ho}(2023)}]{zhang23}
{Zhang}, L., \& {Ho}, L.~C. 2023, \apjl, 953, L9,
  \dodoi{10.3847/2041-8213/acea73}

\end{thebibliography}
\bibliographystyle{aasjournal}



\end{CJK*}
\end{document}